\documentclass[twoside,a4paper]{jpconf_mod}
\usepackage{graphicx,wrapfig,amssymb} %url
\pdfoutput=1

\begin{document}
\title[Experimental prospects at the Large Hadron Collider]{Experimental prospects at the Large Hadron Collider\footnote[99]{In this review, only the detector and physics performance of ATLAS and CMS, the two general-purpose experiments at the Large Hadron Collider, will be addressed. Two other large experiments, namely LHCb~\cite{LHCb} and ALICE~\cite{ALICE}, dedicated to B-physics and heavy ions, respectively, will also operate at the LHC. Finally TOTEM~\cite{TOTEM} and LHCf~\cite{LHCf} will focus on forward physics at the LHC.}}

\author{Daniel Froidevaux$^1$ and Vasiliki A Mitsou$^2$}

\address{$^1$ CERN, PH Department, CH-1211 Geneva 23, Switzerland}
\address{$^2$ Instituto de F\'{i}sica Corpuscular (IFIC), CSIC -- Universitat de Val\`encia, \\ Apartado de Correos 22085, E-46071 Valencia, Spain}

\begin{abstract}
This review focuses on the expected performance of the ATLAS and CMS detectors at the CERN Large Hadron Collider~(LHC), together with some of the highlights of the global commissioning work done in 2008 with basically fully operational detectors. A selection of early physics measurements, expected to be performed with the data taken in 2009/2010 is included for completion, together with a brief reminder of the ultimate physics potential of the LHC.
\end{abstract}

\section{Introduction}%%%%%%%%%%%%%%%%%%%%%%%%%%%%%%%%%%%%%%%%%%%%%%%%%%%%%%%%%%%%%%%%%

The Large Hadron Collider (LHC)~\cite{LHC}, located at CERN, the European Laboratory for Particle Physics, outside Geneva, Switzerland, will hopefully see its first collisions at a centre-of-mass energy of~10~TeV towards the end of~2009. In September~2008, excitement reached its peak after a very successful start-up of the machine with single beams, but an unfortunate incident caused substantial damage in one of the sectors of the machine and delayed the first collisions by about one year. 

The LHC accelerator complex and its experiments will provide a long-awaited and unprecedented tool for fundamental physics research for many years to come. Ultimately, the LHC will provide two proton beams, circulating in opposite directions, at their design energy of 7~TeV each, corresponding to a centre-of-mass energy of~$\sqrt{s} = 14$~TeV), and at a design luminosity of~$\rm10^{34}~cm^{-2}s^{-1}$. The two general-purpose experiments, ATLAS (A Toroidal LHC ApparatuS)~\cite{ATLAS} and CMS (Compact Muon Solenoid)~\cite{CMS}, share a long and common history: they were conceived in the early~90's at the same time as a world-wide Research and Development effort began, focused on the numerous technical challenges associated with operation in the harsh environment of the~LHC. They were approved in~1994, the detailed design of the various parts of the apparatus was completed and reviewed between~1997 and~2002, the construction proceeded between 1997 and 2007, and the integration and installation at~CERN was finally completed during the summer of~2008. The experiments were therefore ready just in time for first beams in the fall of~2008. Next year is therefore hopefully going to see the culmination of the work of thousands of people across the world over almost twenty years.

\section{ATLAS/CMS: from design to reality}%%%%%%%%%%%%%%%%%%%%%%%%%%%%%%%%%%%%%%%%%%%%

In the following, the basic requirements which drove the design of the~ATLAS and~CMS detectors, the technological choices adopted by the collaborations, and the performance expected for the measurement and identification of the major physics objects are reviewed. 

\subsection{An unprecedented scale and complexity}

The constraints of the unprecedented energy and luminosity of the LHC~accelerator have led to detectors, which are colossal in many aspects and of considerable complexity in many others:
\begin{description}
\item[Size of detectors] With a volume of~$\rm20\,000~m^3$, ATLAS is one of the largest accelerator-based experiments built so far, while CMS is one of the heaviest, weighing~12\,500~tons. Both detectors feature 70~to~80~million pixel readout channels near the interaction point. The active silicon of the CMS~tracker covers~$\rm200\,m^2$, whereas the ATLAS liquid argon~(LAr) electromagnetic~(EM) calorimeter contains a total of~175\,000 readout channels. For the detection and momentum measurement of muons, the experiments are equipped with approximately one million channels over an area of~$\rm10\,000\,m^{2}$ of muon chambers. The most sensitive components of the front-end electronics are distributed across the whole detector volume and all the materials and active elements must remain operational after having received the very large radiation doses expected over the lifetime of the experiments. The very selective trigger/DAQ system provides a rejection of approximately~$10^7$ in real time. The real and simulated data are handled by large-scale offline software and world-wide connected computing facilities.

\item[Time-scale] About 25~years will have passed from the first conceptual studies of the~LHC (Lausanne 1984~\cite{Lausanne}) to solid physics results, which will mark the transition of the high-energy frontier from the Tevatron to the LHC, presumably in~2010. For many scientists, this period will have represented a major chunk of their professional life and career.

\item[Size of collaboration] The number of authors of the first physics papers will have jumped from~$\sim600$ for the~LEP and Tevatron experiments to two to three thousand for~ATLAS and~CMS. The amount of information circulated and meetings occurring regularly is staggering but appears to be necessary for the collaborations to manage their own work and evolution.
\end{description}

\begin{table}[ht]
\caption{\label{tab:xsections}For a few physics processes among those expected to be the most abundantly produced at the LHC, expected numbers of events recorded by ATLAS and CMS for an integrated luminosity of 1~fb$^{-1}$ per experiment~\cite{annual}.} 
\begin{center}
\lineup
\begin{tabular}{*{2}{l}}
\br                              
Physics processes & Number of events per fb$^{-1}$\cr 
\mr
QCD jets with $E_{\rm T}> 150$~GeV & $10^6$ (for 10\% of trigger bandwidth)\cr
${\rm W \rightarrow \mu\nu}$ & $7\times10^6$ \cr 
${\rm Z \rightarrow \mu\mu}$ & $1.1\times10^6$ \cr 
${\rm t\bar{t} \rightarrow e/\mu} + X $& $1.6\times10^5$ \cr 
$\rm \tilde{g}\tilde{g}$ production ($m_{\tilde{\rm g}}\approx1$~TeV) & $10^2$ to $10^3$ \cr
\br
\end{tabular}
\end{center}
\end{table}

 \newpage
\subsection{The experimental challenge at the LHC}

\begin{wrapfigure}[33]{r}{0.51\textwidth}
% \vspace{-15mm}
  \begin{center}
    \includegraphics[width=0.48\textwidth]{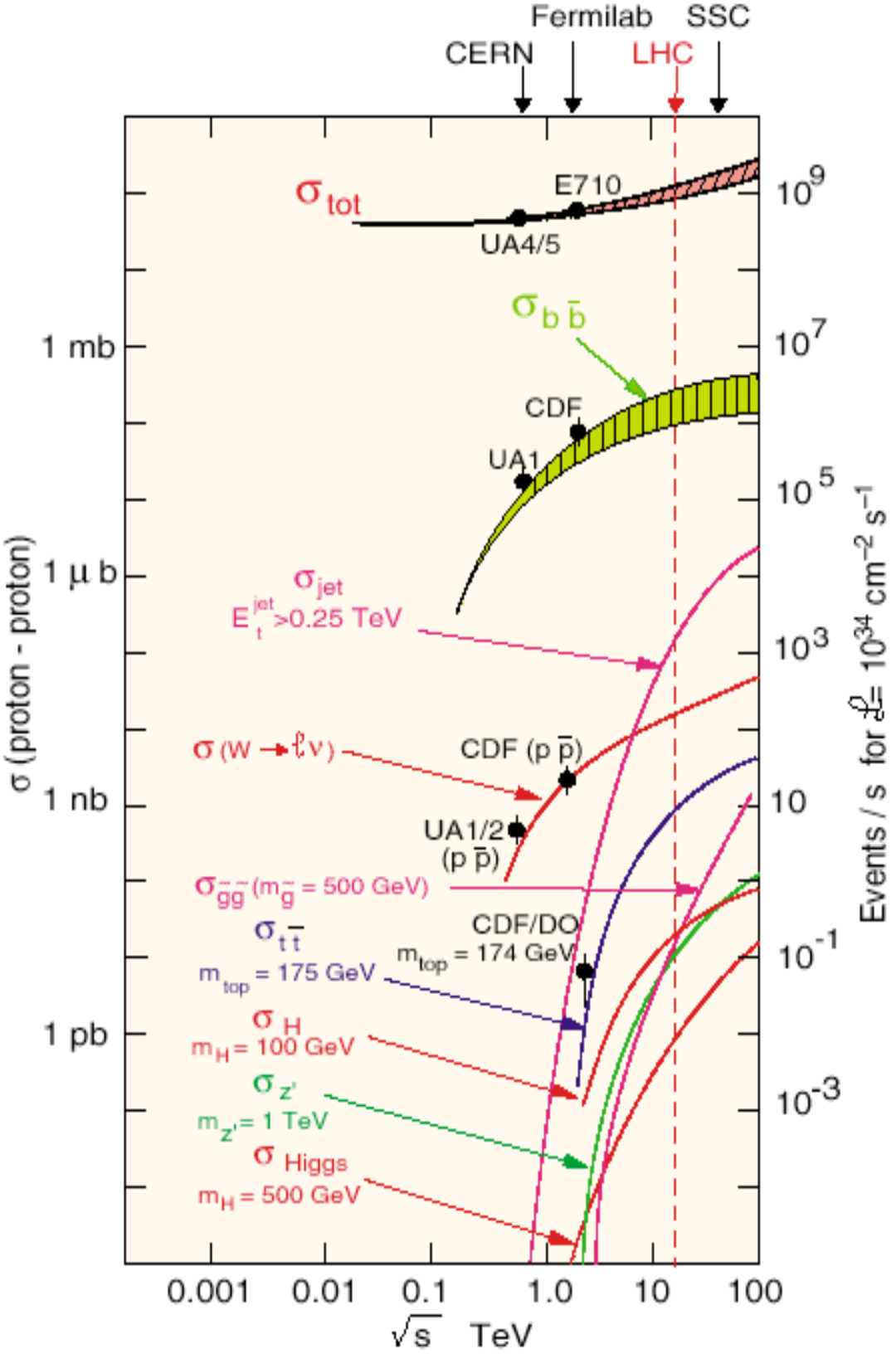}
  \end{center}
  \caption{\label{fig:xsections}Cross sections and event rates at design luminosity for hard scattering processes as a function of the centre-of-mass energy $\sqrt{s}$~\cite{altarelli}.}
\end{wrapfigure}

The LHC represents the next major step in the high-energy frontier beyond the Fermilab Tevatron (proton-antiproton collisions at a centre-of-mass energy of $\sim2$~TeV). The high design luminosity is required because of the small cross-sections expected for many of the benchmark processes used to optimise the design of the general-purpose detectors over the past 15~years or so. The cross-sections and expected event rates for processes such as leading Standard Model~(SM) processes, Higgs-boson production, supersymmetry, etc., are shown in~figure~\ref{fig:xsections} as a function of $\sqrt{s}$ and in~table~\ref{tab:xsections} as expected rates at $\sqrt{s}$~=~14~TeV for an integrated luminosity of~1~fb$^{-1}$ per experiment. To achieve such luminosities and minimise the impact of inelastic collisions occurring simultaneously in the detectors (a phenomenon known as pile-up), the LHC~beam crossings are 25~ns apart in time, resulting in 23~inelastic interactions on average per crossing at design luminosity.

In order to cope with the potentially overwhelming SM processes, high rejection power is needed with optimal efficiency for rare channels such as many of the Higgs-boson decays. In the most extreme case, a selection power of $10^{-14}-10^{-15}$ is required for Higgs-boson discovery when small signal rates are compared to the total interaction rate. This represents an increase of many orders of magnitude over the selection rates achieved at the Tevatron to-date. 

The QCD jet production cross-section is overwhelmingly dominant over all electroweak processes: this means that lepton signatures are most often essential in extracting rare processes from the background at the~LHC. It is for this very reason that both excellent measurements and superb identification capabilities of both electrons and muons have been among the primordial requirements of both ATLAS and CMS since the very early design days. Additional signatures have also driven the global design to a large extent, e.g.~missing transverse energy,~$E{\rm _T^{miss}}$, typically indicating the presence of high-$p_{\rm T}$ non-interacting particles such as neutrinos, or secondary vertices, typically indicating the presence of heavy-flavour hadrons usually embedded in high-$p_{\rm T}$ $b/c$-jets. 

In addition to the physics challenge itself, the LHC~experiments have had to face an even more daunting challenge from the very beginning, namely that of the unprecedented radiation levels expected at design luminosity. At the LHC, the primary source of radiation at design luminosity comes from collisions at the interaction point. In the inner detector, charged hadrons from inelastic proton-proton interactions dominate the radiation backgrounds at small radii, while other sources, such as neutrons, become more important at larger radii, typically above~50~cm.

%\begin{figure}[ht]
%\includegraphics[width=0.45\textwidth]{xsections}\hspace{0.05\textwidth}%
%\begin{minipage}[b]{0.5\textwidth}\caption{\\}
%\end{minipage}
%\end{figure}

In this context, the generic features required of the ATLAS and CMS detectors and embodied in their global design are the following:
\begin{description}
\item[Detector survival for at least 10~years of operation]
 The high radiation levels will damage the detectors themselves as well as their electronics components. Near the interaction point, in particular, many years of research and development have been spent on selecting the materials, designing front-end electronics, and certifying assembled parts in nuclear reactors and under high ionising radiation, before construction could begin in earnest. The problem of detector survival pervades the whole experimental area at the~LHC, because of the high neutron fluences produced in the forward regions where most of the beam energy is dissipated.

\item[Precise timing and fast detector response]
The bunch-crossing interval of~25~ns provides a direct requirement in this area: in 25~ns, even a relativistic particle only travels a distance of~7.5~m across the detector, whereas the corresponding electronics signals only travel a distance of~5~m out of the typically $\sim100$-metre-long readout cables. These numbers only hint at the massive effort which had to be put into accurate and absolutely reliable timing and control signals throughout the front-end electronics and the trigger system. The source of these timing signals is a single precision clock synchronised with the LHC~beams.
 
The speed of the detector response is another requirement which can be met in certain technologies but not in all of them. A slower response means increased sensitivity to out-of-time interactions from preceding or subsequent interactions, thereby compounding the problem of pile-up mentioned above.

\item[Fine spatial granularity]
A very fine spatial granularity is another ingredient to minimise the impact of pile-up events since an interesting signal in a certain region of the detector occurring at a certain time will only be distorted significantly by other particles if they produce both pile-up in time and in space. Considerations such as these, as well as the occupancy in the core of high-$p_{\rm T}$ jets have driven some of the choices for the granularity of the tracking detectors and electromagnetic calorimetry.

\item[Identification of extremely rare events]
As already mentioned, excellent lepton identification should be achieved with high efficiency to extract rare signals from the huge QCD~backgrounds. For instance, the electron-to-jet ratio at the~LHC is~$\sim10^{-5}$ at~$p_{\rm T} \sim 20$~GeV, a factor $\sim 100$~times worse than at the Tevatron.

The online rejection to be achieved in real time is~$\sim 10^7$ and huge data volumes have to be recorded to permanent storage, typically~$\sim 10^9$~events of 1~Mbyte~size per year.
\end{description}

\subsection{Main design choices}\label{sc:design}

The size of the ATLAS and CMS experiments are directly related to the energies of the  particles produced. The calorimeters must absorb the energy of electrons of several~TeV (approximately 30~radiation lengths,~$X_0$, corresponding for instance to~18~cm of lead are required) and of pions of similar energies (approximately 11~interaction lengths,~$\lambda$, corresponding for instance to~2~m of iron are required). The muon spectrometers and tracking detectors must measure precisely the momenta muons up to several~TeV, either inside or outside the calorimeters: $BL^2$,~the product of the magnetic field strength,~$B$, and of the distance,~$L$, traversed by the muon, is a key factor in this case which must be carefully optimised.

The choice of the magnet system has shaped the experiments in a major way. A strong magnetic field is required to measure the momenta and directions of charged particles near the interaction vertex: this is usually provided by a solenoid, bending the particles in the plane transverse to the proton beams. A strong magnetic field is also required to trigger and if possible measure precisely muon momenta outside the calorimeters (muons are the only charged particles not absorbed in calorimeter absorbers). 

In the case of ATLAS, two separate magnet systems have been chosen: a small 2~T~solenoid for the tracker and huge toroids with large~$BL^2$ for the muon spectrometer. This option offers a large acceptance in polar angle for the muons (thanks to the toroidal field, which extends to large pseudorapidities) and excellent muon momentum resolution even without using the tracker information. The main drawback of this choice is that it has resulted in a very complex and expensive design and construction project to successfully produce a very large-scale toroid magnet system.

In the case of CMS, on the other hand, the collaboration has opted for a very elegant solution with one large 4~T~solenoid with an instrumented iron return yoke. This solution provides both excellent momentum resolution using the tracker and adequate triggering capabilities outside the calorimeter with muon stations embedded in the iron return yoke. This design also leads to a more compact experiment. The muon performance is however limited for stand-alone muon measurements (and for the trigger at very high luminosities) and at small angles to the beam, and the sheer size of the coil allows for only limited space for the calorimeters inside it.

\begin{figure}[ht]
\begin{center}
\includegraphics[width=0.95\textwidth]{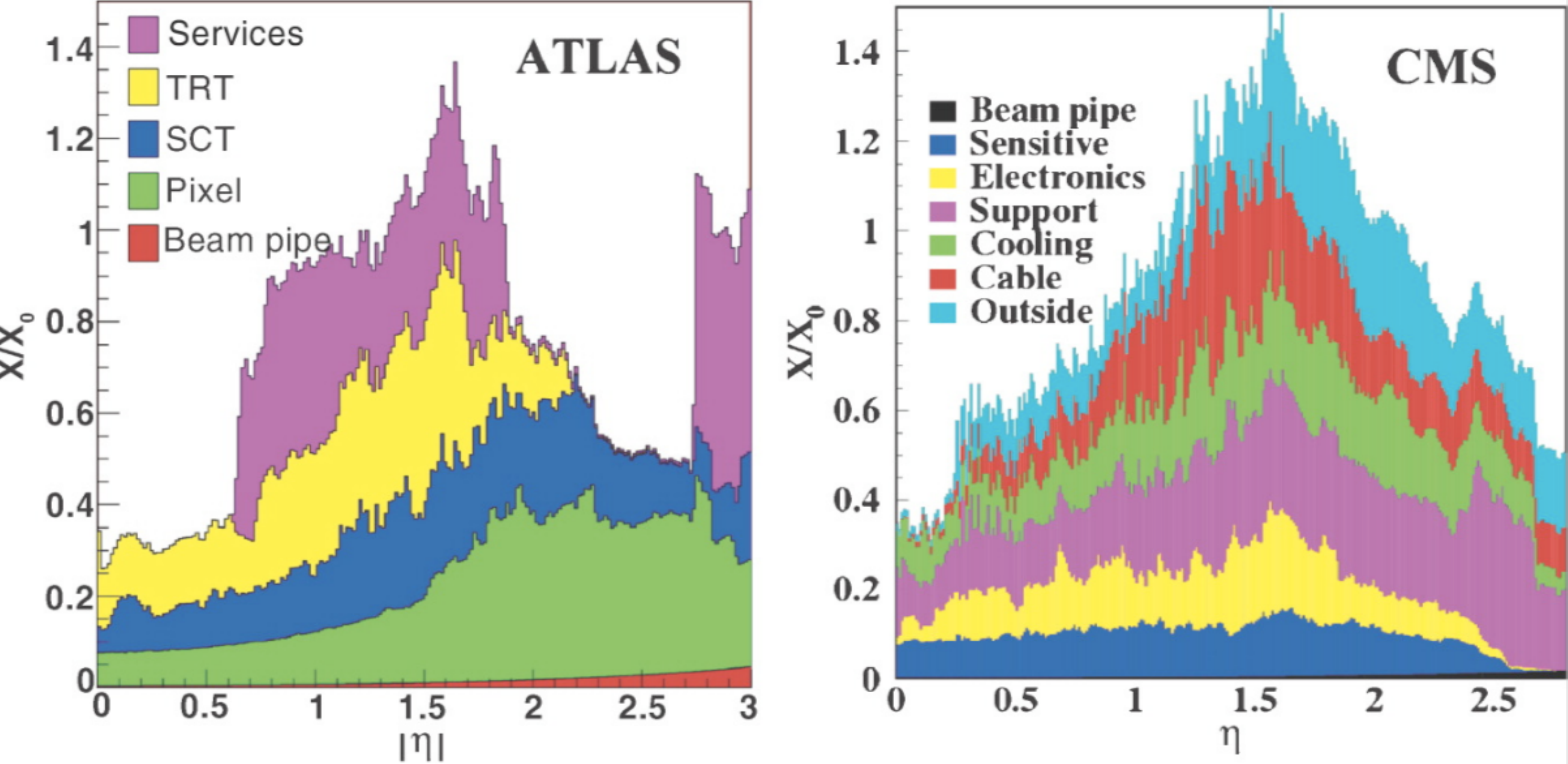}
\end{center}
\caption{\label{fig:material}Distribution of amount of material in the volume of the ATLAS (left) and CMS (right) trackers, expressed as fractional radiation length $X/X_0$ versus pseudorapidity $\eta$. These plots do not include additional material just in front of the electromagnetic calorimeters, which is quite large in ATLAS (LAr cryostat and, for the barrel, solenoid coil) and much less in CMS (front part of crystal mechanics)~\cite{annual}.}
\end{figure}

One major concern has emerged over the years as the detailed design of the trackers developed and converged towards a complete definition of all materials inside the envelope of the electromagnetic calorimeters: the material budget of the tracking systems is far superior to that of any of the preceding collider experiments (LEP or Tevatron), and it has increased by a factor of~2 to~2.5 from~1994 (approval of the experiments) to now (operational detectors). The amount of material present in the trackers of~ATLAS and~CMS is shown in~figure~\ref{fig:material}. Even at small pseudorapidities, the amount of material is $3-4$~times larger than in the LEP experiments and it represents much more than one radiation length at its peak. Electrons therefore lose between~25\% and~70\% of their energy and 20\%~to~65\% of photons convert into $\rm e^+e^-$~pairs before reaching the electromagnetic calorimeter. This large and inhomogeneous amount of material has to be known to a precision of~$\sim1\%~X_0$ for the most accurate measurements planned at the~LHC, for example the measurement of~$m_{\rm W}$ to less than~10~MeV.

At the LHC, which is essentially a gluon-gluon collider, the unambiguous identification and precise measurement of leptons is the key to many areas of physics. Electrons are relatively easy to measure precisely in electromagnetic calorimeters but are very hard to identify, as already discussed above. Muons, in contrast, are relatively easy to identify behind the calorimeters, but very hard to measure accurately at high energies. These aspects have also shaped to a large extent the global design and technology choices of the two experiments.

The electromagnetic calorimetry of ATLAS and CMS is based on very different technologies: ATLAS uses a LAr~sampling calorimeter with good energy resolution and excellent lateral and longitudinal segmentation, while~CMS has chosen $\rm PbWO_4$~scintillating crystals with excellent energy resolution and lateral segmentation, but without any longitudinal segmentation. Broadly speaking, signals from Higgs-boson decays, such as $\rm H\rightarrow\gamma\gamma$ or $\rm H\rightarrow ZZ^* \rightarrow 4e$, should appear as narrow peaks (intrinsically narrower in~CMS) above a fairly pure background from the same final state (intrinsically purer in terms of fakes in ATLAS).

\section{Expected performance}

\subsection{Inner trackers}

\begin{figure}[ht]
\begin{minipage}[b]{0.485\textwidth}
\includegraphics[width=\textwidth]{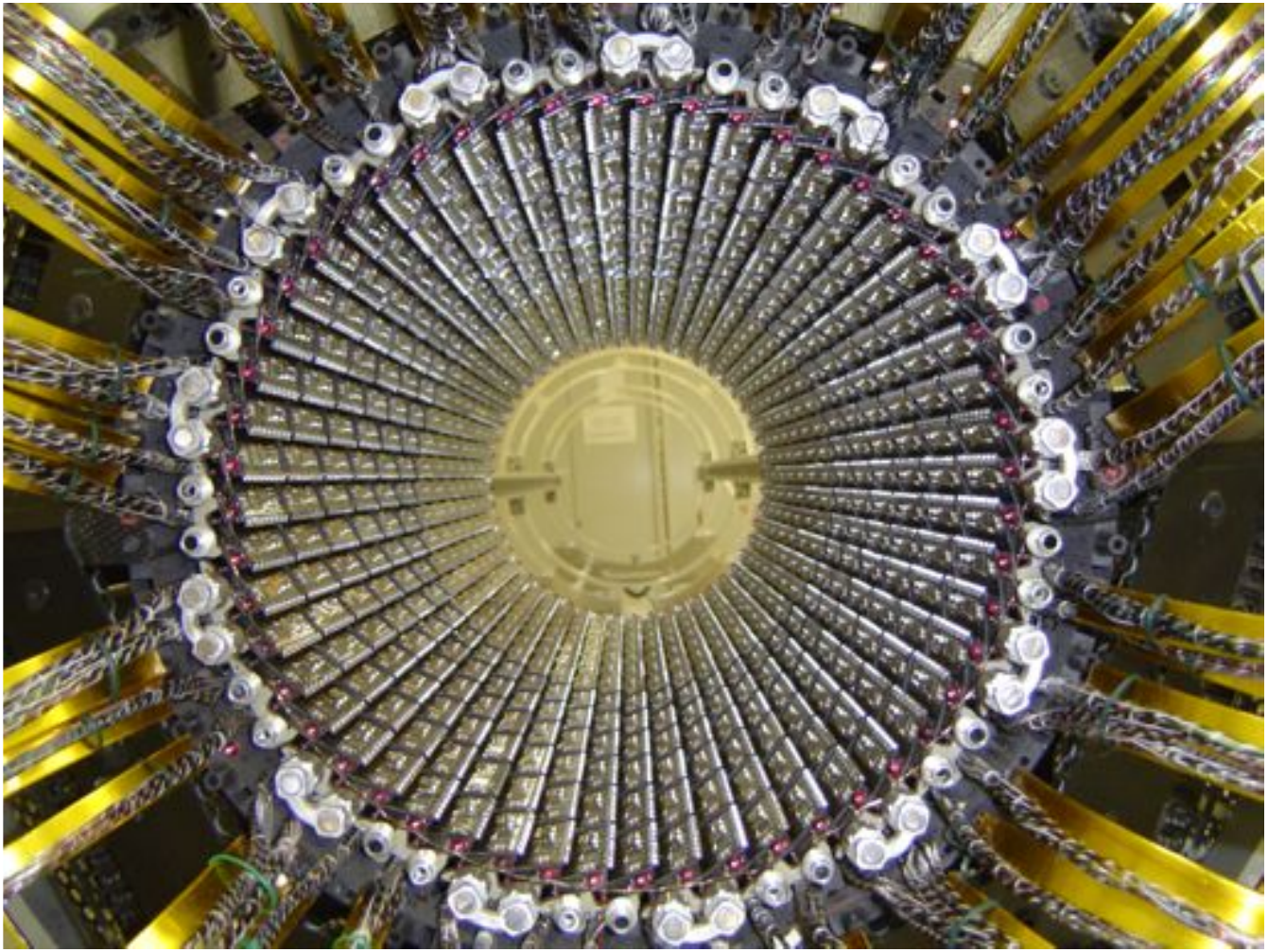}
\caption{\label{fig:ATLASpixel}A picture of the ATLAS Pixel innermost layer taken during the pixel barrel integration in~2006.}
\end{minipage}\hspace{0.05\textwidth}%
\begin{minipage}[b]{0.485\textwidth}
\includegraphics[width=\textwidth]{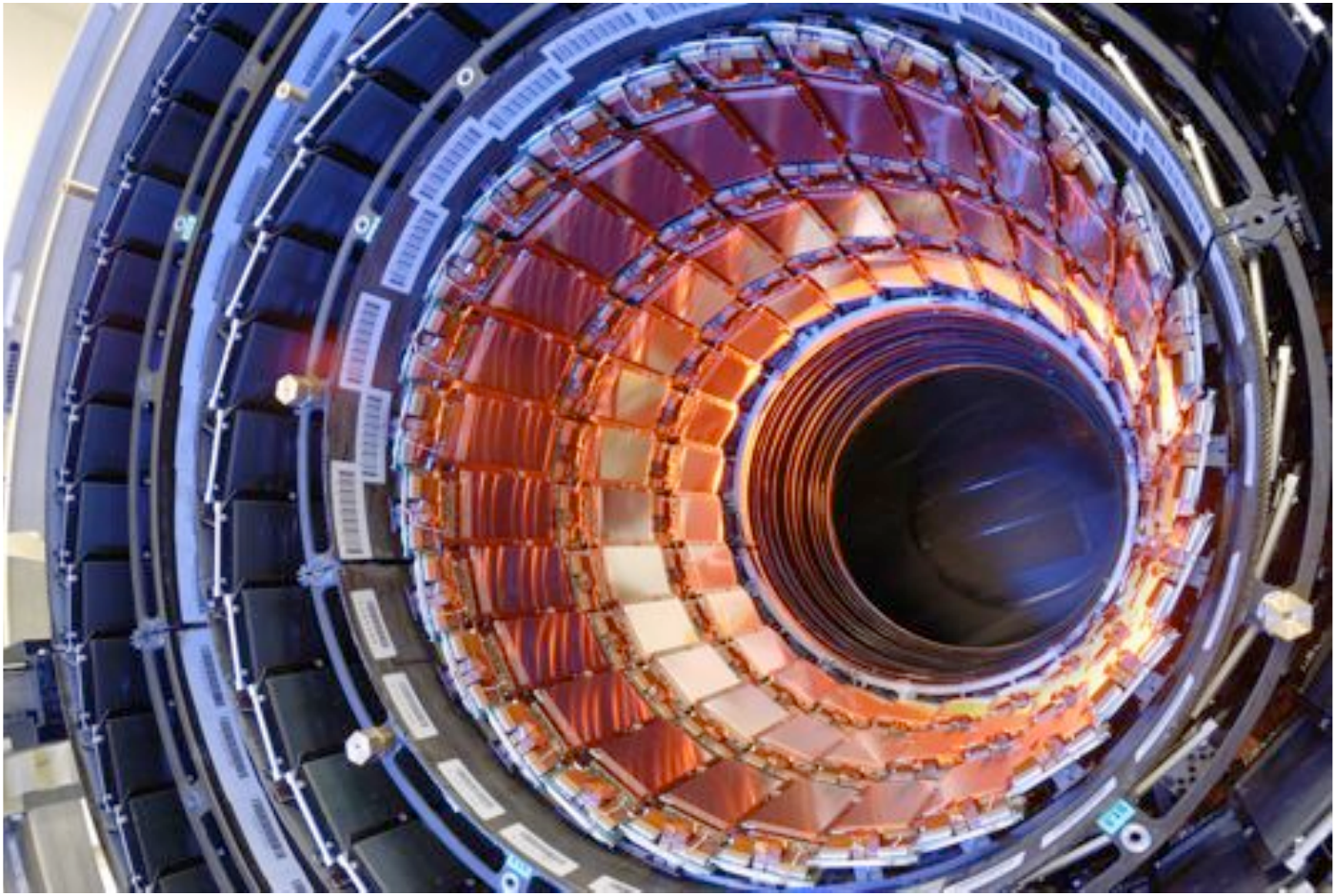}
\caption{\label{fig:CMS_tracker}A spectacular view of the interior of one-half of the~CMS  inner barrel tracker, showing many of the silicon sensors.}
\end{minipage} 
\end{figure}

ATLAS and CMS have designed tracker systems, which provide the same geometrical coverage ($|\eta|<2.4$--2.5) and are similar near the interaction vertex, but which differ considerably in terms of the technological choices made at larger radii. The most challenging part of each project, namely the pixel detectors (see~figure~\ref{fig:ATLASpixel}), which surround the interaction point, provide a set of three measurements per primary track at small radii. 

At intermediate radii, both trackers contain a set of thin and fine-pitched silicon-strip detectors, providing at least eight (resp.~six) measurements per track for~ATLAS (resp.~CMS), in both the bending plane and along the $z$-axis through stereo layers.

CMS has extended the silicon-strip technology all the way to the outermost radii, thus providing eight additional measurements per track, with coarser-pitch detectors to cover this large volume. In ATLAS, in contrast, at larger radii, an average of~35~measurements per track in the bending plane are provided by thin straw-tube detectors operating with a Xenon-based gas mixture. The straws are embedded in fibre or foil radiators providing electron identification through absorption of the X-rays from transition radiation, which is produced by highly relativistic particles traversing the multiple interfaces.

Table~\ref{tab:tracker} compares the main performance parameters of the~ATLAS and CMS~trackers, as obtained from extensive simulation studies performed over the years and benchmarked using detailed test-beam measurements of production modules wherever possible. The unprecedentedly large amount of material present in the trackers is reflected in the overall reconstruction efficiency for charged pions of low transverse momentum, which is only slightly above~80\%, as opposed to the value of~97\% obtained for muons of the same transverse momentum. The electron track reconstruction efficiency is even more affected by the tracker material, and the numbers shown in~table~\ref{tab:tracker} for electrons of 5~GeV transverse momentum are only indicative of the performance expected, as the efficiency obtained depends strongly on the criteria used to define a reasonably well-measured electron track. The somewhat lower efficiencies obtained in the case of~CMS are probably due to the higher magnetic field, which enhances effects arising from interactions in the detector material.

\begin{table}[ht]
\caption{\label{tab:tracker}Main performance characteristics of the ATLAS and CMS trackers: examples of typical reconstruction efficiencies, momentum resolutions, and transverse and longitudinal impact parameter~(i.p.) resolutions are given for various particle types, transverse momenta, and pseudorapidities~\cite{annual}.} 
\begin{center}
\lineup
\begin{tabular}{*{3}{l}}
\br                              
Tracking properties & ATLAS & CMS \cr
\mr
Reconstruction efficiency for pions with $p_{\rm T} = 1$~GeV & 84.0\% & 80.0\% \cr
Reconstruction efficiency for electrons with $p_{\rm T} = 5$~GeV & 90.0\% & 85.0\% \cr
Momentum resolution at $p_{\rm T} = 100$~GeV and $\eta\approx0$ & 3.8\% & 1.5\% \cr
Momentum resolution at $p_{\rm T} = 100$~GeV and $\eta\approx2.5$ & 11\% & 7\% \cr
Transverse i.p.\ resolution at $p_{\rm T} = 1000$~GeV and $\eta\approx0$ & 11~$\rm\mu m$ & 9~$\rm\mu m$ \cr
Transverse i.p.\ resolution at $p_{\rm T} = 1000$~GeV and $\eta\approx2.5$ & 11~$\rm\mu m$ & 11~$\rm\mu m$ \cr
Longitudinal i.p.\ resolution at $p_{\rm T} = 1000$~GeV and $\eta\approx0$ & 90~$\rm\mu m$ & 22--42~$\rm\mu m$ \cr
Longitudinal i.p.\ resolution at $p_{\rm T} = 1000$~GeV and $\eta\approx2.5$ & 190~$\rm\mu m$ & 70~$\rm\mu m$ \cr
\br
\end{tabular}
\end{center}
\end{table}

The performance of the CMS tracker is undoubtedly superior to that of~ATLAS in terms of momentum resolution and longitudinal impact parameter resolution. THe overall vertexing and $b$-tagging performances of the two experiments are similar. The impact of the tracker material and of the $B$-field is however already visible on the individual efficiencies, perhaps more so in~CMS. The~ATLAS and~CMS trackers are expected to deliver performances close to those specified at the time of their conceptual design fifteen years ago, despite the harsh environment in which they will operate for many years and the difficulty of the many technical challenges encountered along the way. In contrast to most of the other systems in the two experiments, however, they will not survive nor deliver the required performance if the LHC~luminosity is upgraded to~$\rm 10^{35}~cm^{-2}s^{-1}$. The~ATLAS and~CMS trackers will therefore have to be replaced by detectors with finer granularity and with an order of magnitude higher resistance to radiation to meet the challenges of the~SLHC. 

\subsection{Electromagnetic calorimetry}

The ATLAS and CMS electromagnetic~(EM) calorimeters are each divided into a barrel part covering approximately $|\eta| < 1.5$ and two end-caps covering $1.4 < |\eta| < 2.5$ (resp.~3.0) for~ATLAS (for~CMS). The fiducial coverage of these calorimeters is without appreciable cracks, except perhaps in the transition region between the barrel and end-cap cryostats in the case of ATLAS, where the measurement accuracy is degraded because of large energy losses in the material in front of the active EM~calorimeter. The excellent uniformity of coverage is due to the accordion-shaped electrodes and absorbers of the ATLAS Pb/LAr~sampling calorimeter. In~CMS, this is obtained by the rotation of the CMS PbWO$_4$~crystals away from a purely projective arrangement. The total thickness of the EM~calorimeters varies from~$24~X_0$ to~$35~X_0$. This depth is sufficient to contain EM~showers at the highest energies (a few~TeV) and thereby preserve the energy resolution, in particular the constant term, which is dominant above a few hundred GeV.

\begin{figure}[ht]
\begin{center}
\includegraphics[width=0.47\textwidth]{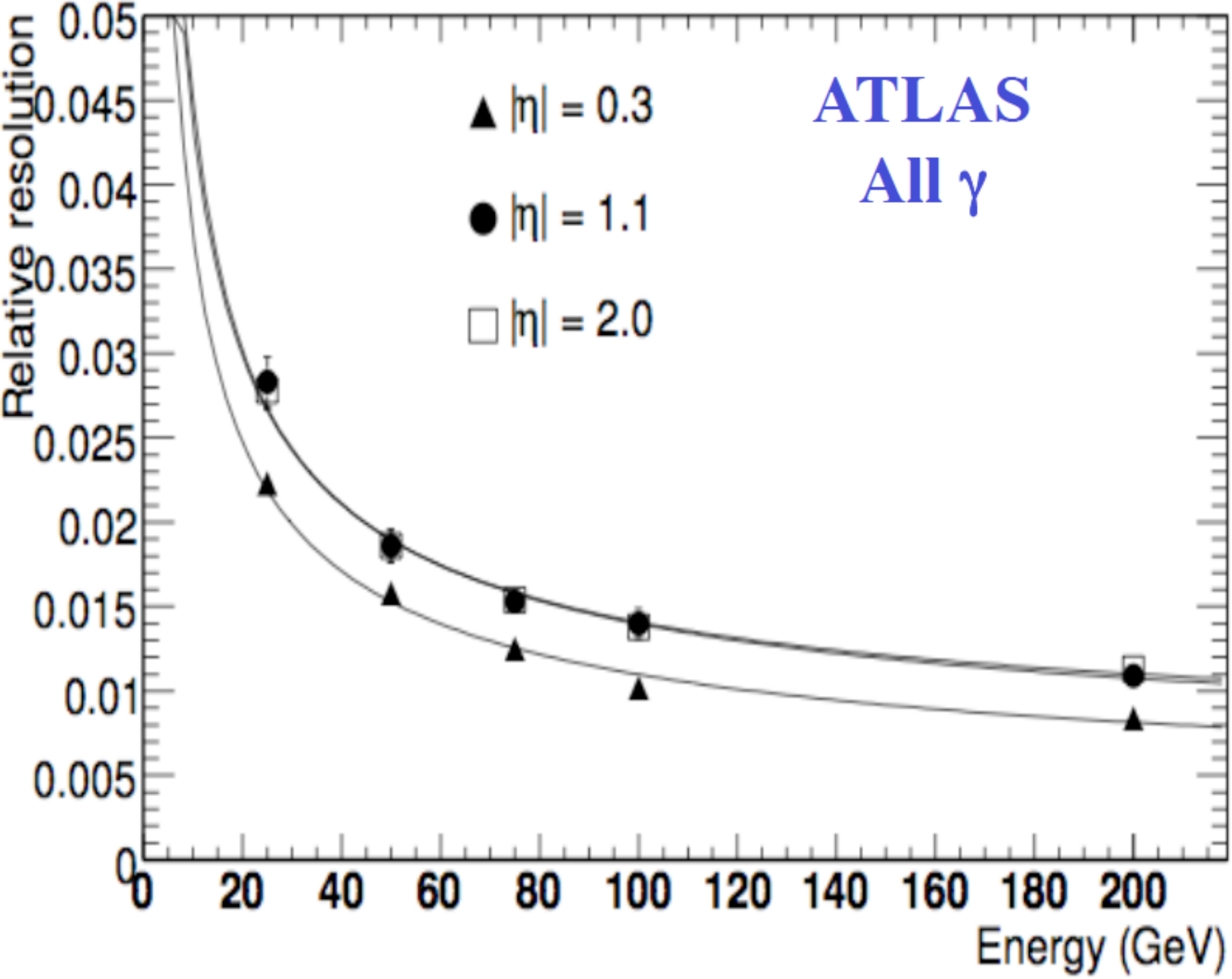}
\hspace{0.04\textwidth}
\includegraphics[width=0.45\textwidth]{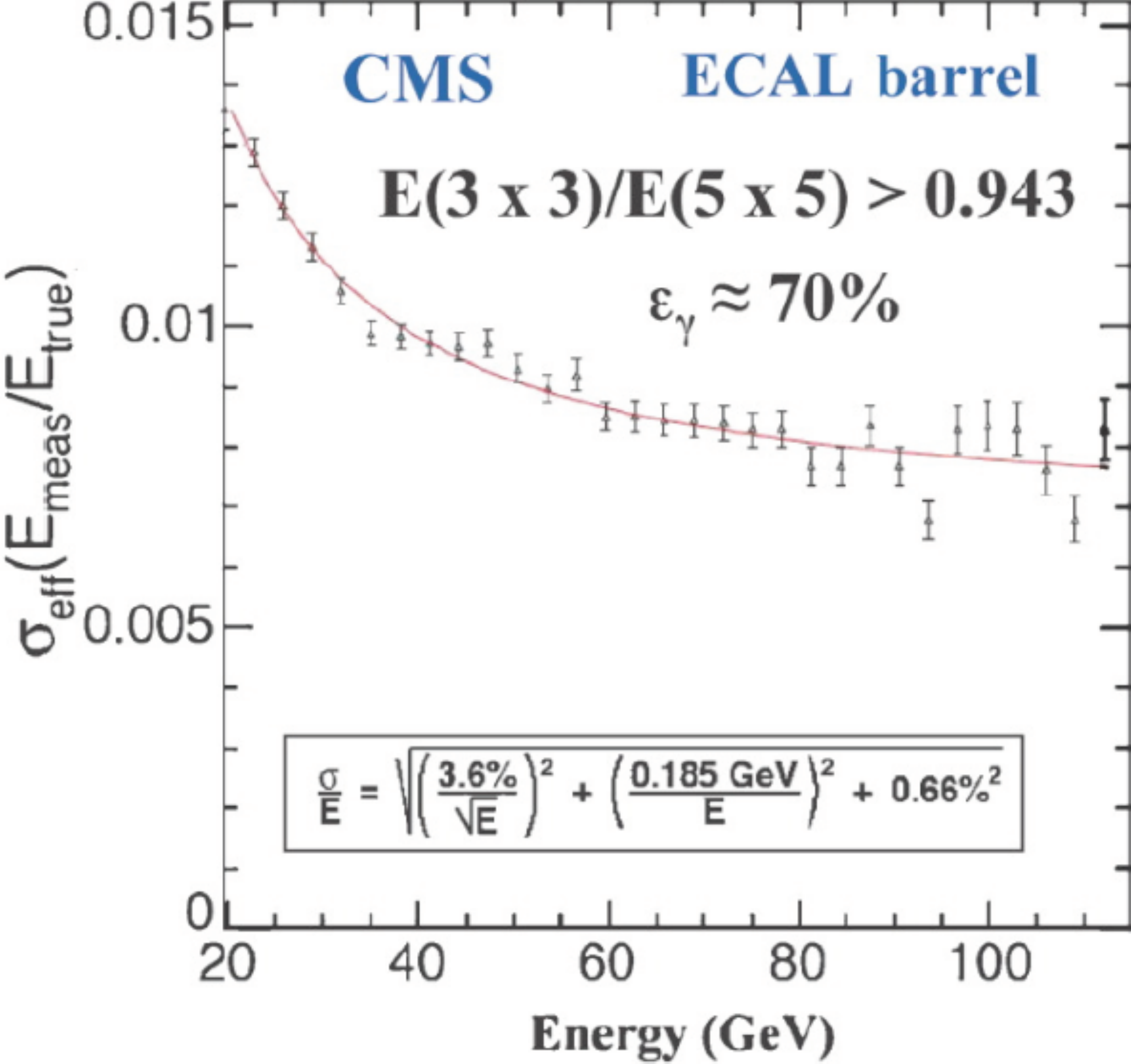}
\end{center}
\caption{\label{fig:perf_photons}For ATLAS (left) and CMS (right), expected relative precision on the energy measurement of photons as a function of their energy. Also shown are fits to the stochastic, noise, and local constant terms of the calorimeter resolution.}
\end{figure}

Figure~\ref{fig:perf_photons} shows an example of the expected precision with which ATLAS and CMS will perform photon energy measurements. For~ATLAS, results are shown for unconverted and converted photons together and for a few typical values of~$\eta$, over the energy range from~20 to~200~GeV. For CMS, the results are shown for predominantly unconverted photons in the barrel crystal calorimeter and in the energy range from 20 to 110~GeV. For a photon energy of~100~GeV, the ATLAS energy resolution varies between~1.0\% and~1.4\% over the full $\eta$-range. These numbers increase respectively to~1.2\% and 1.6\% if the expected global constant term of~0.7\% is included. The overall expected CMS energy resolution in the barrel crystal calorimeter is 0.75\% for the approximately 70\%~of well-measured photons at that energy before including the expected global constant term of~0.5\%. This example shows that the intrinsic resolution of the CMS crystal calorimeter is harder to obtain with the large amount of tracker material in front of the EM~calorimeter and in the 4~T~magnetic field: between~20\% and~60\% of photons in the barrel calorimeter acceptance convert before reaching the front face of the crystals.

\begin{figure}[ht]
\begin{center}
\includegraphics[width=0.47\textwidth]{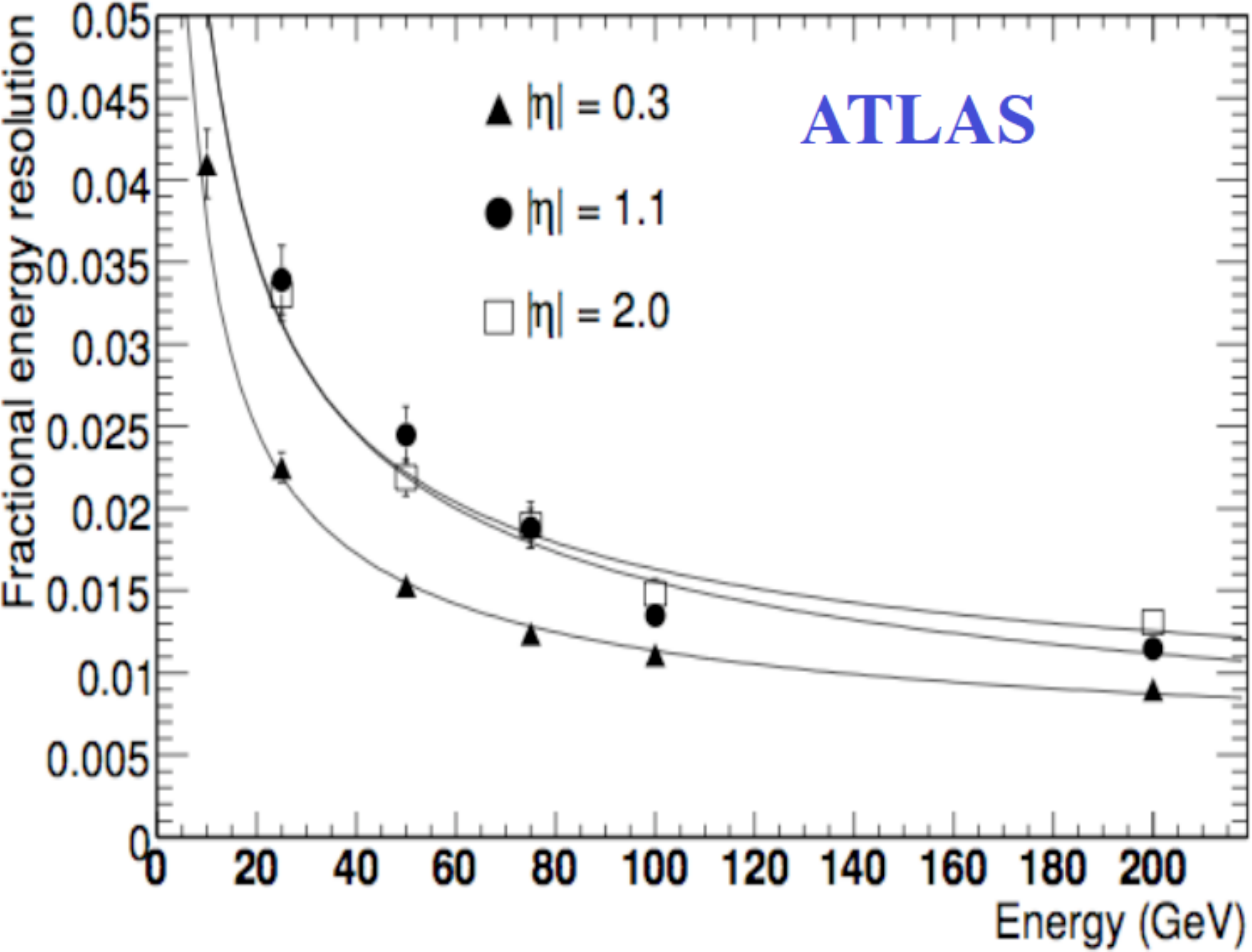}
\hspace{0.04\textwidth}
\includegraphics[width=0.46\textwidth]{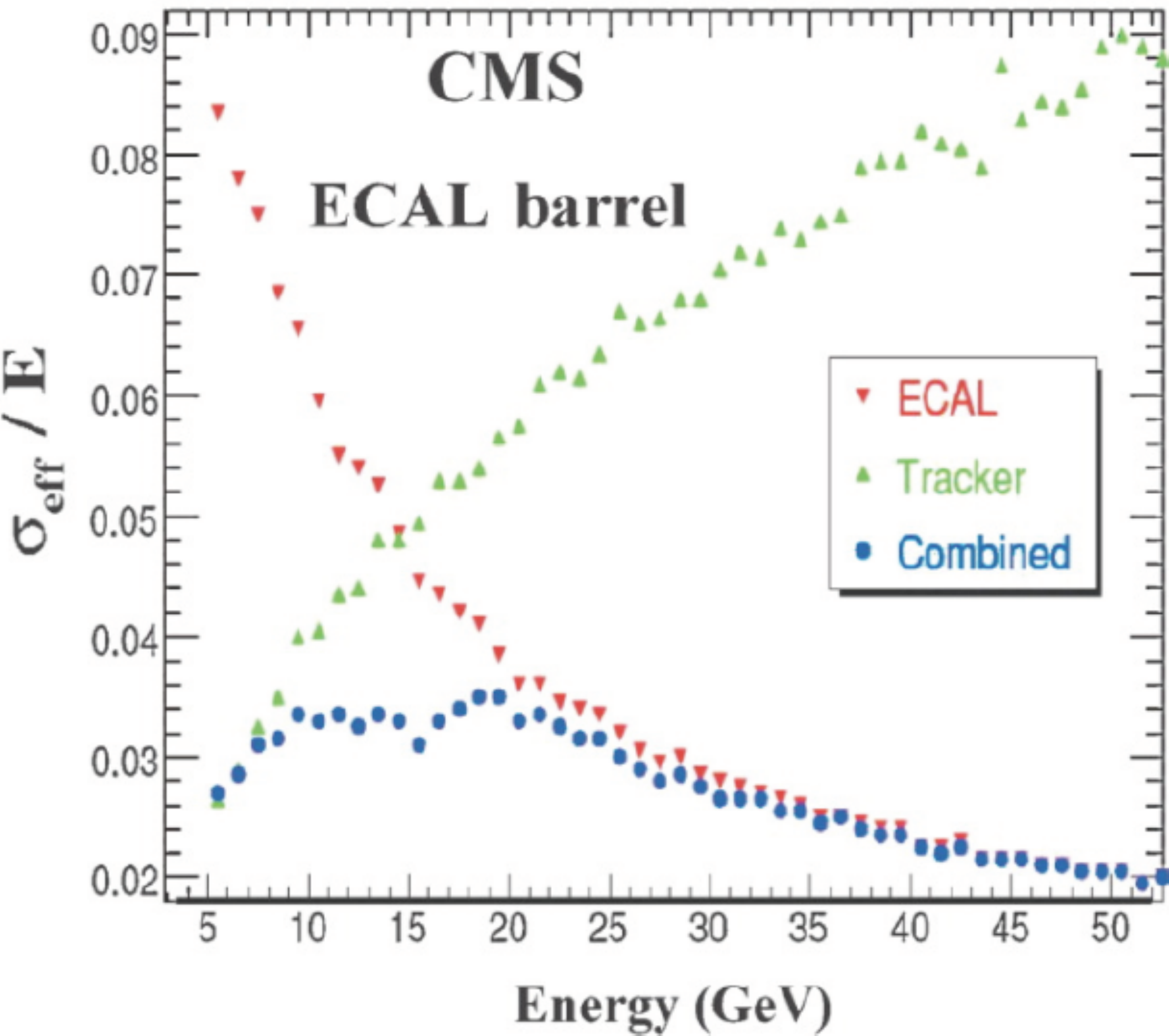}
\end{center}
\caption{\label{fig:perf_electrons}For ATLAS (left) and CMS (right), expected relative precision on the energy measurement of electrons as a function of energy. The resolutions are shown for ATLAS for a few typical values of~$\eta$ together with fits to the stochastic and local constant terms of the calorimeter resolution. For CMS, the combined effective resolution at low energy is shown over the acceptance of the barrel crystal calorimeter, together with the individual contributions from the tracker and the EM~calorimeter.}
\end{figure}

Similarly, figure~\ref{fig:perf_electrons} shows an example of the expected precision with which ATLAS and CMS will perform electron energy measurements. For ATLAS, results are shown for a few typical values of~$\eta$ over the energy range from~10 to~200~GeV. The energy of the electrons is always collected in a $3\times7$~cell matrix, which is wider in the bending direction to collect as efficiently as possible the bremsstrahlung photons while preserving the linearity and low sensitivity to pile-up and noise. For CMS, the effective resolution (r.m.s.\ spread) is shown for the barrel crystal calorimeter and in the most difficult low-energy range from~5 to~50~GeV. Refined algorithms are used, in both the tracker and the calorimeter, to recover as much as possible the bremsstrahlung tails and to restore thus most of the excellent intrinsic resolution of the crystal calorimeter. For electrons of 50~GeV, the ATLAS energy resolution varies between~1.5\% and~1.8\% over the $\eta$-range, without any specific requirements on the performance of the tracker at the moment. In contrast, the CMS effective resolution is estimated to be~2\%, demonstrating that it is more difficult in the actual experiment to reconstruct electrons than photons with a performance in terms of efficiency and energy resolution similar to the intrinsic one obtained in test-beam.

\subsection{Hadronic calorimetry}

The ATLAS and CMS hadronic calorimeters display a number of significant differences in their design parameters. The strong constraints imposed by the CMS~solenoid have resulted in a barrel hadronic calorimeter with an insufficient absorption length of~$7.2~\lambda$ at~$\eta = 0$ for the complete calorimeter including the crystals before the coil, so a tail catcher has been added around the coil to complement the calorimetry and to provide better protection against punch-through to the muon system. Consequently, the sampling fraction in the CMS hadronic calorimetry is approximately three times worse than that in the ATLAS hadronic calorimetry, which explains to a great extent the better hadronic resolution expected in ATLAS.

\begin{figure}[ht]
\begin{center}
\includegraphics[width=0.44\textwidth]{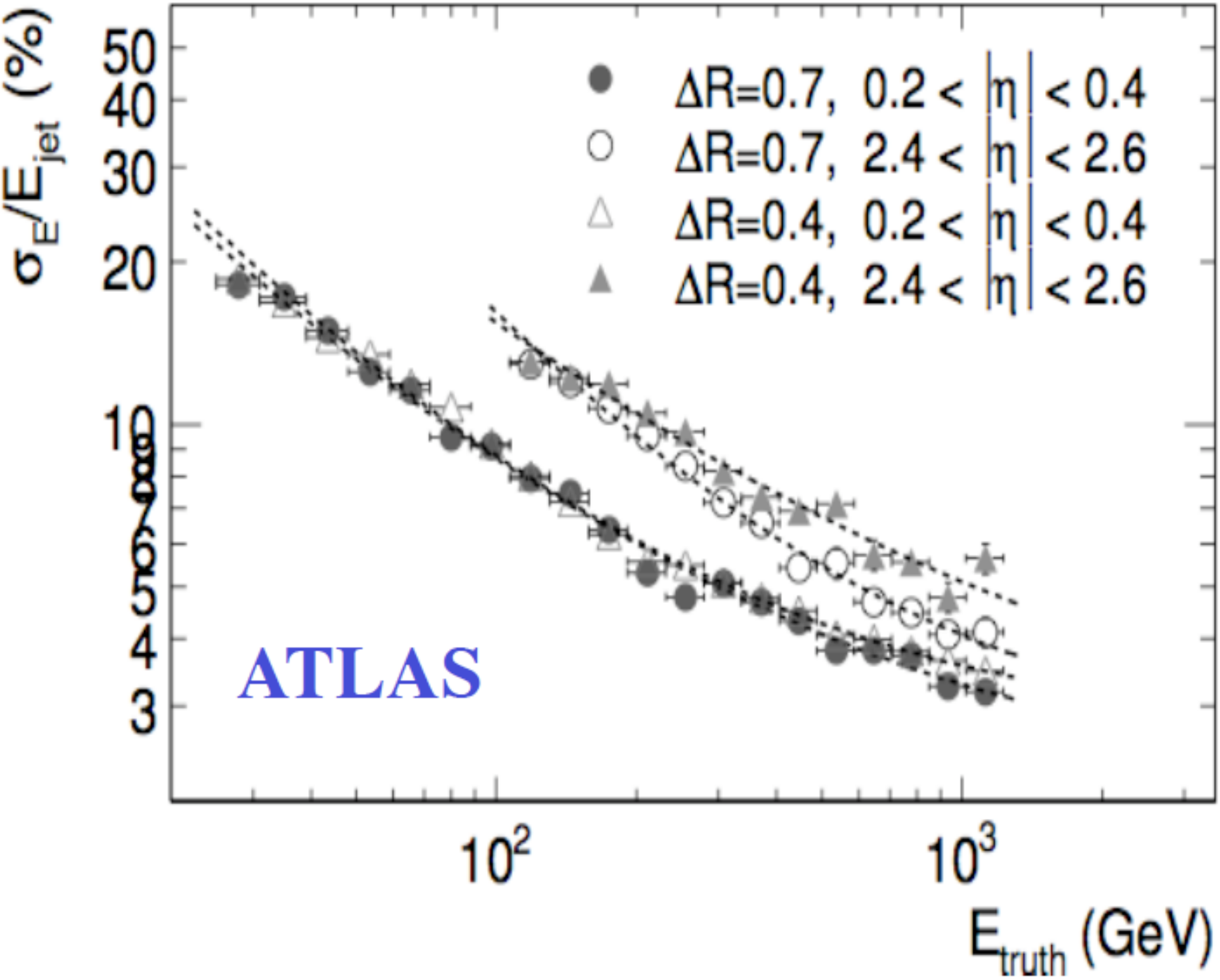}
\hspace{0.04\textwidth}
\includegraphics[width=0.46\textwidth]{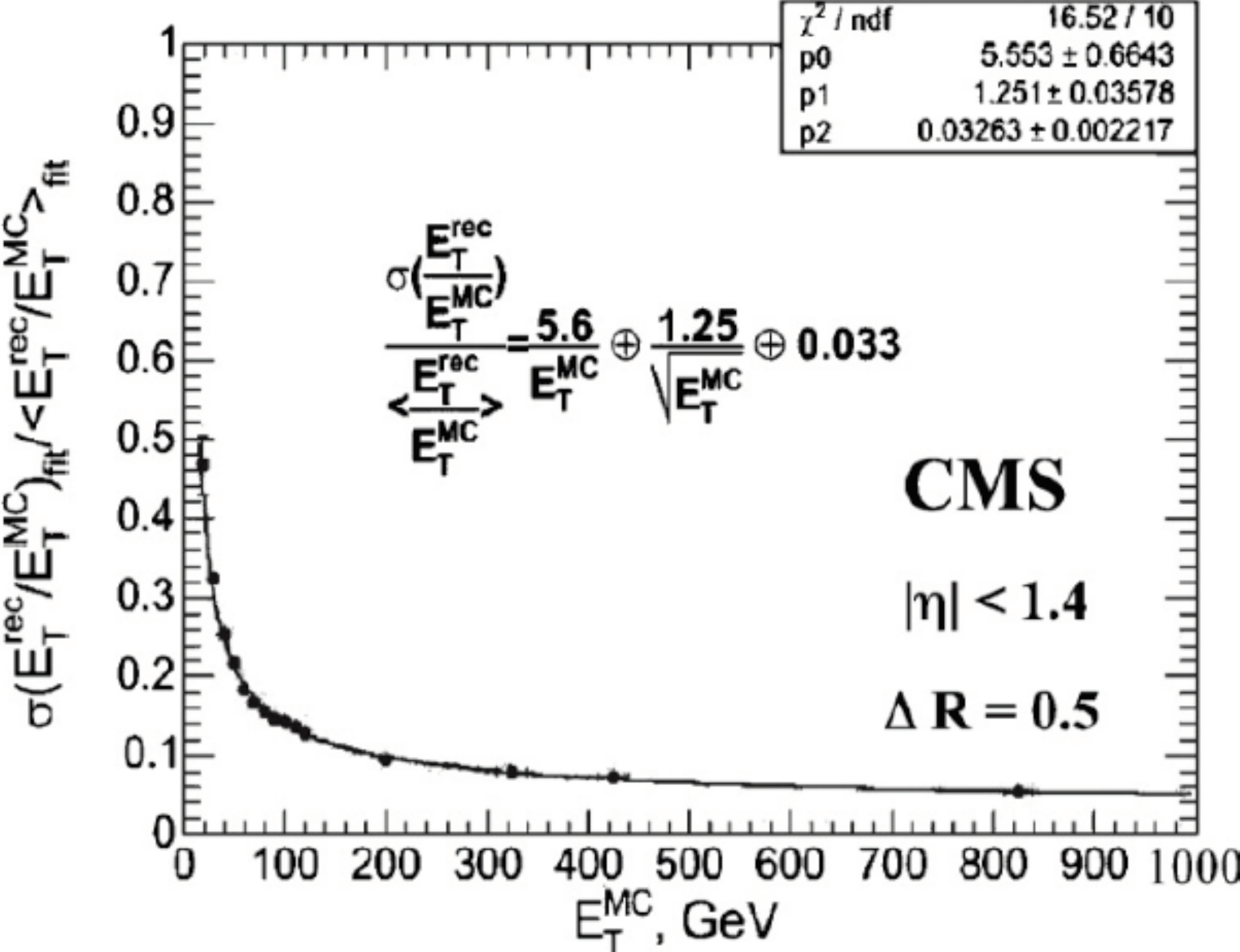}
\end{center}
\caption{\label{fig:perf_jets}Expected relative precision on the measurement of the reconstructed energy of QCD jets as a function of the jet energy and for two $\eta$-ranges for ATLAS~(left), and as a function of~$E_{\rm T}^{\rm MC}$, where  $E_{\rm T}^{\rm MC}$ is the jet transverse energy, in the barrel region for~CMS~(right).}
\end{figure}

The expected performance for reconstructing hadronic jets is shown in~figure~\ref{fig:perf_jets}. Jets are found using a simple cone algorithm with a fixed size in pseudorapidity-azimuth space, so that all cells within a distance $\Delta R=\sqrt{\Delta\eta^2+\Delta\phi^2}$ of a seed cell are included. For ATLAS, the jet energy resolution is depicted for two $\eta$-regions and two different cone sizes over an energy range from~30 to~1000~GeV. The jet energy resolution is shown using a sophisticated weighting technique inspired by the work done in the H1~collaboration. For~CMS, the jet energy resolution is shown for a cone size of~$\Delta R = 0.5$ and for~$|\eta| < 1.4$, over a transverse energy range from~15 to~800~GeV. For completeness, figure~\ref{fig:perf_jets} also displays the results of fits to the stochastic, noise, and local constant terms of the calorimeter resolutions. For hadronic jets of typically 100~GeV~energy, characteristic of jets from W-boson decay, the ATLAS energy resolution in the barrel region is approximately~8\%, whereas the CMS energy resolution is approximately~14\%. 

\begin{figure}[ht]
\begin{center}
\includegraphics[width=0.44\textwidth]{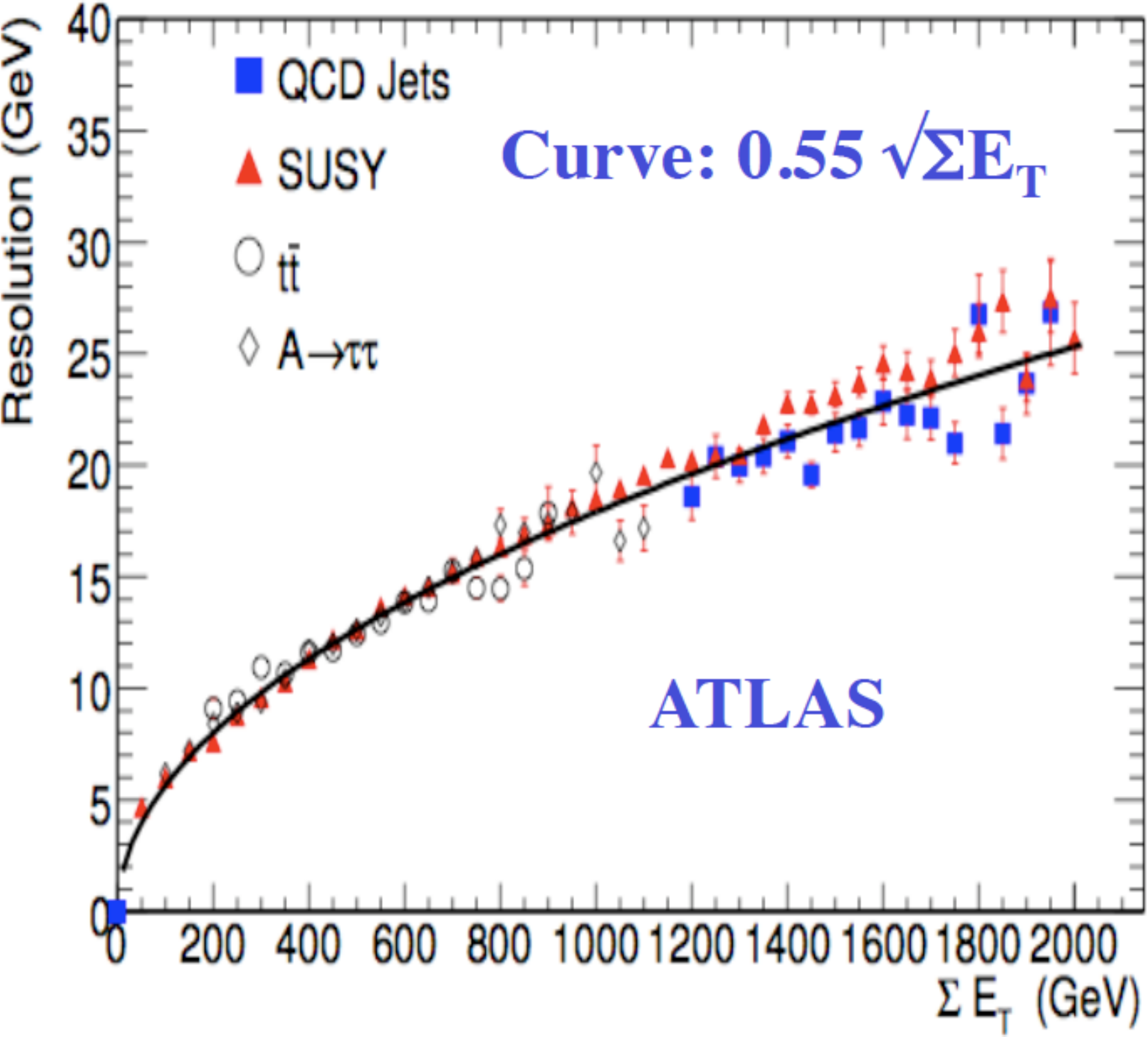}
\hspace{0.04\textwidth}
\includegraphics[width=0.43\textwidth]{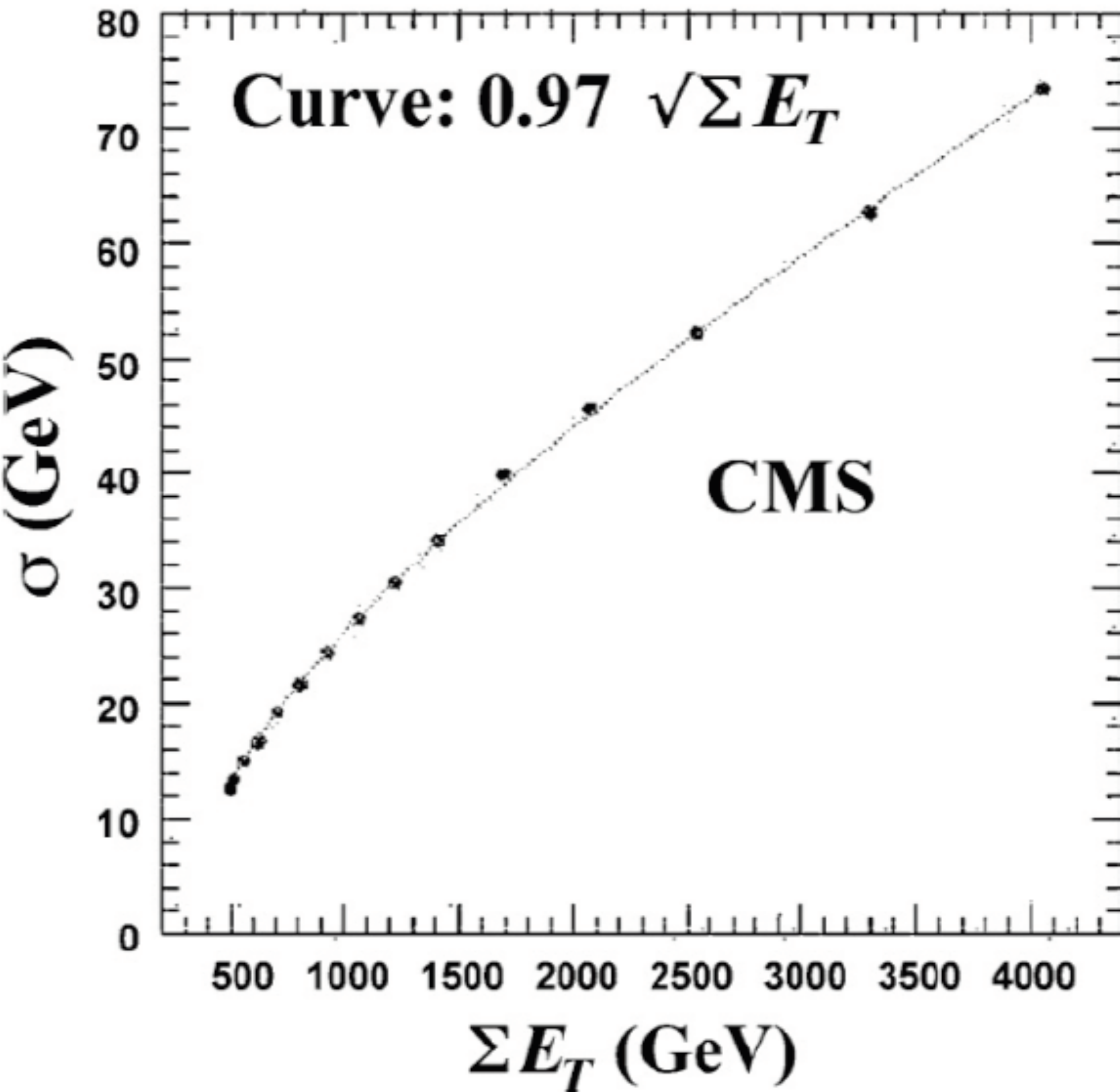}
\end{center}
\caption{\label{fig:perf_MET}For ATLAS (left) and CMS (right), expected precision on the measurement of the missing transverse energy as a function of the total transverse energy, $\Sigma E_{\rm T}$, measured in the event.}
\end{figure}

Finally, figure~\ref{fig:perf_MET} illustrates an important aspect of the overall calorimeter performance, namely, the expected precision with which the missing transverse energy in the event can be measured in each experiment as a function of the total transverse energy deposited in the calorimeter. The results are expressed as variances of Gaussian fits to the $(x,\,y)$~components of the missing transverse energy vector, for events from high-$p_{\rm T}$ jet production and for $\rm A\rightarrow\tau\tau$ decays in ATLAS. For CMS, where the distributions are non-Gaussian, the results are expressed as the~r.m.s.\ of the same distributions for events from high-$p_{\rm T}$ jet production. For transverse momenta of the hard-scattering process ranging from~70 to~700~GeV, the reconstructed $\Sigma E_{\rm T}$ ranges from approximately~500~GeV to approximately~2~TeV. The difference in performance between~ATLAS and~CMS is a direct consequence of the difference in performance already expected for the jet energy resolution.

One word about neutrinos in hadron colliders: since most of the energy of the colliding protons escapes down the beam pipe, one can only use the energy-momentum balance in the transverse plane. Therefore, concepts such as $E{\rm _T^{miss}}$, missing transverse momentum and mass are often used (the only non-measurable component is~$E_z^{\rm miss}$). These observables allow the complete reconstruction of certain topologies with neutrinos, e.g.\ $\rm W\rightarrow\ell\nu$ and more importantly $\rm Z/H\rightarrow\tau\tau\rightarrow\ell\nu_{\ell}\nu_{\tau}h\nu_{\tau}$. The experiments must therefore be quite hermetic, so that the transverse energy flow can be fully measured with reasonable accuracy and no neutrino escapes undetected.

\subsection{Muon spectrometers}

The ability to trigger and reconstruct muons at the highest luminosities of the LHC was incorporated into the design of the ATLAS and CMS detectors from the beginning. In fact, the concepts chosen by the two experiments for measuring muon momenta have shaped the two detectors more than any other physics consideration and, as discussed in~Section~\ref{sc:design}, the choice of magnet was motivated by the measurement method of muons with TeV-scale momenta. 

The ATLAS toroidal magnetic field provides a momentum resolution that is essentially independent of pseudorapidity up to a value of~2.7. ATLAS has opted for a high-resolution, stand-alone measurement independent of the rest of the subdetectors, resulting in a large volume with low material density over which the muon measurement takes place. The muon system consists of three large superconducting air-core toroid magnets, which are instrumented with different types of chambers to provide two necessary functions, namely, high-precision tracking and triggering. In the central region ($|\eta| < 1.0$), covered by a large barrel magnet consisting of eight coils surrounding the HCAL, tracks are measured in chambers arranged in three cylindrical layers around the beam axis. In the end-cap region, $1.4<\eta<2.7$, muon tracks are bent in two smaller end-cap magnets each inserted into one end of the barrel toroid. The transition region, $1.0<\eta<1.4$, is less straightforward because here the barrel and end-cap fields overlap, thus partially reducing the bending power. To keep a uniform resolution in this region, tracking chambers, which allow corrections for the change in the magnetic field, are strategically placed.

\begin{figure}[ht]
\includegraphics[width=0.43\textwidth]{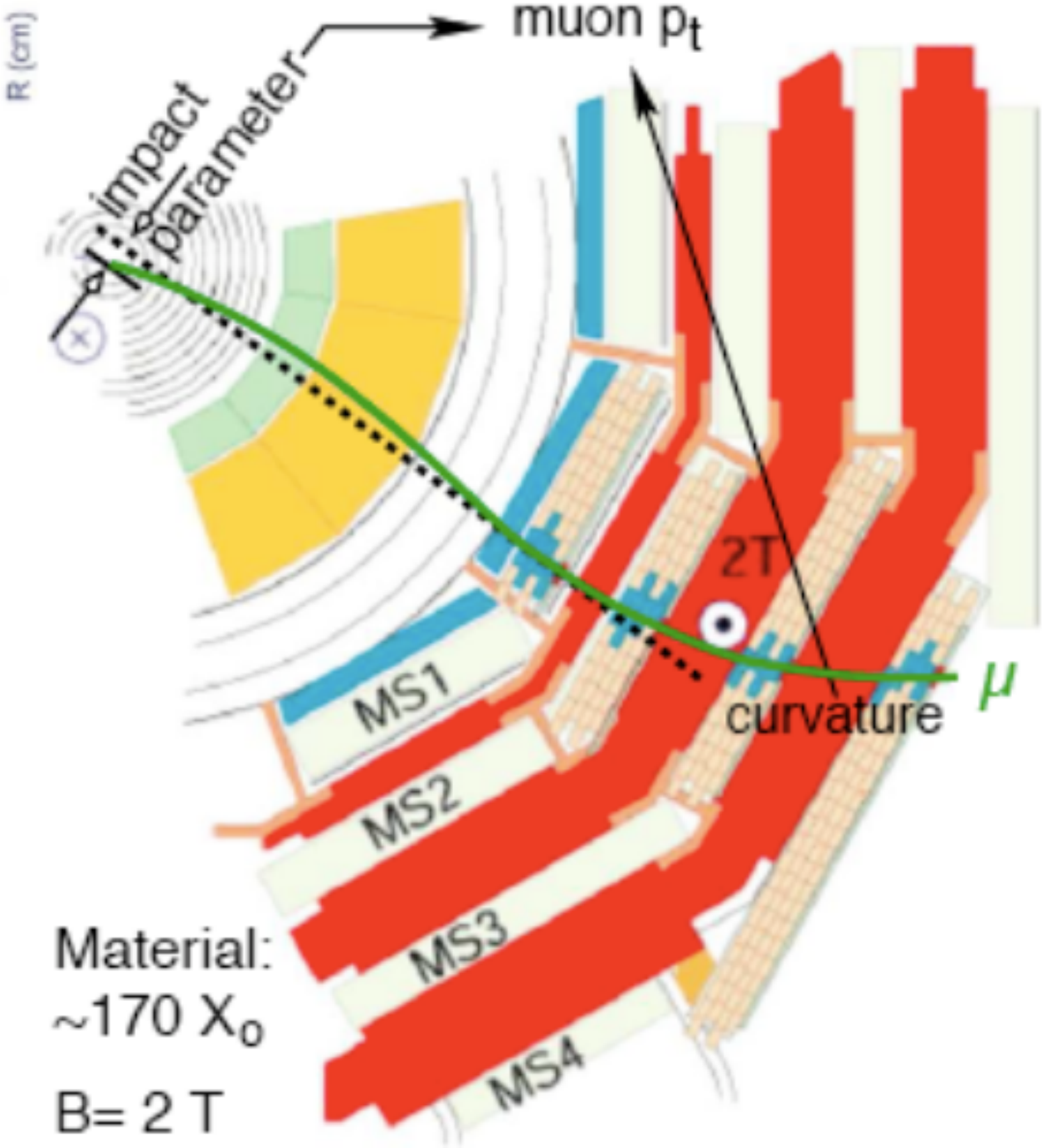}\hspace{0.05\textwidth}%
\begin{minipage}[b]{0.5\textwidth}\caption{\label{fig:CMSmuon}A schematic view of a slice of the CMS detector in the transverse plane. The magnetised iron is in red/dark-grey, and the track trajectory (green solid line) is measured in the inner tracker (hatched circles inside the blue-yellow/light-grey calorimeter) and in the pink/light-grey (blue/dark-gey those hit by the track) muon chambers. Charged particle tracks are bent one way in the tracker and then the other way outside it. \\}
\end{minipage}
\end{figure}

CMS, on the other hand, uses the concept of a compact detector, which therefore needed a high magnetic field to provide enough bending power for the muon measurement. The CMS solenoidal field bends muon tracks in the transverse plane, as displayed in~figure~\ref{fig:CMSmuon}, effectively adding another point ---the primary vertex position, which is expected to be known with high accuracy--- to the muon track. The muon system consists of chambers installed between the iron slabs that provide the return yoke for the magnetic field of the solenoid. In the barrel region, the detectors are arranged in cylinders interleaved with the iron yoke. In the end-caps, the chambers are arranged in four disks perpendicular to the beam and in concentric rings, three rings in the innermost station and two in the others. 

Table~\ref{tab:muon} lists the expected resolutions on the muon momentum measurement for the two spectrometers. The expected near independence of the resolution from the pseudorapidity in ATLAS and the degradation of the resolution at high~$\eta$ in CMS are clearly visible. The resolution of the combined measurement in the barrel region is better in CMS owing to the higher resolution of the measurement in the tracking system, whereas the reverse is true in the end-cap region for high muon momenta owing to the better coverage of the ATLAS toroidal system at large pseudorapidities. 

\begin{table}[ht]
\caption{\label{tab:muon}Summary of the expected combined and stand-alone (in parenthesis) muon momentum resolution at two typical pseudorapidity values (averaged over azimuth) and for various momentum values~\cite{annual}.} 
\begin{center}
\lineup
\begin{tabular}{*{3}{l}}
\br                              
Momentum and pseudorapidity & ATLAS & CMS \cr
\mr
  $p=10$~GeV and $\eta\approx 0$ & 1.4\% (3.9\%) & 0.8\% (8\%) \cr
  $p=10$~GeV and $\eta\approx 2$ & 2.4\% (6.4\%) & 2.0\% (11\%) \cr
 $p=100$~GeV and $\eta\approx 0$ & 2.6\% (3.1\%) & 1.2\% (9\%) \cr
 $p=100$~GeV and $\eta\approx 2$ & 2.1\% (3.1\%) & 1.7\% (18\%) \cr
$p=1000$~GeV and $\eta\approx 0$ & 10.4\% (10.5\%) & 4.5\% (13\%) \cr
$p=1000$~GeV and $\eta\approx 2$ & 4.4\% (4.6\%) & 7.0\% (35\%) \cr
\br
\end{tabular}
\end{center}
\end{table}

To recapitulate, the actual muon spectrometer performances match that expected from the original designs. The CMS muon spectrometer provides superior combined momentum resolution in the central region, but the stand-alone resolution and trigger at very high luminosities is limited due to multiple scattering in the iron. The overall resolution is degraded in the forward regions ($|\eta| > 2.0$), where the solenoid bending power becomes insufficient. The ATLAS muon spectrometer exhibits excellent stand-alone capabilities and coverage in an open geometry. However the complicated geometry and field configuration results in large fluctuations in acceptance and performance over the full potential $\eta\times\phi$~coverage. The CMS muon performance is driven by its tracker, so it is better than ATLAS at~$\eta \approx 0$, while the ATLAS muon stand-alone performance is excellent over the whole pseudorapidity range~($|\eta| < 2.7$).

\section{Experience with cosmic rays and first beams}%%%%%%%%%%%%%%%%%%%%%%%%%%%%%%%%%%

\begin{figure}[ht]
\includegraphics[width=0.55\textwidth]{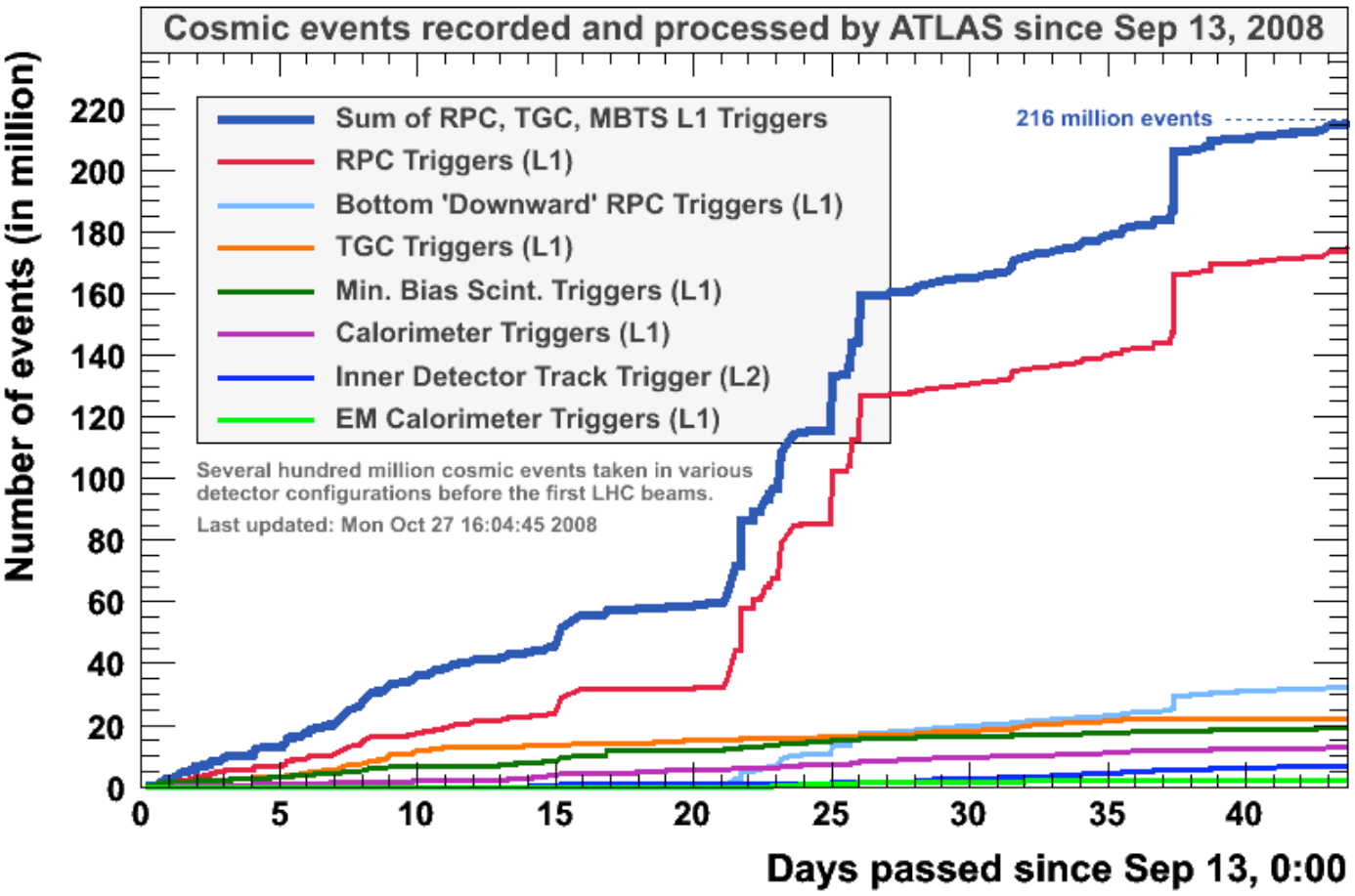}\hspace{0.05\textwidth}%
\begin{minipage}[b]{0.4\textwidth}\caption{\label{fig:cosmics}Integrated cosmic data rate for ATLAS versus time after September $\rm13^{th}$, 2008, i.e., after the single-beam runs. \\}
\end{minipage}
\end{figure}

During the last two years (2007 and 2008), the focus of the ATLAS and CMS commissioning efforts has evolved from single detector operation to combined running and integration. Several combined tests with cosmic rays have been scheduled to integrate detector, trigger and data acquisition into one global setup for each group of sub-systems (calorimeters, muon detectors, inner detector and magnets). These tests have been very helpful in the commissioning of the subsystems, in the integration of controls and of the trigger and data acquisition. The amount of cosmics data collected by the ATLAS detector versus time is displayed in~figure~\ref{fig:cosmics}. 

A central role in the comprehensive pre-collision commissioning campaign was played by the Magnet Test and Cosmic Challenge (MTCC) for the CMS experiment. The objective of the MTCC was the recording, offline reconstruction, and display of cosmic muons in the four subsystems of CMS (tracker, ECAL, HCAL and muon detector) with the magnet operating at its full strength of 4~T. During the test, $25\cdot10^6$ cosmic triggered events were recorded with the principal subdetectors active, of which $15\cdot10^6$ events have a stable field of $\geq3.8$~T. Data-taking efficiency reached over 90\% for extended periods. Several thousand of these events correspond to the `4-detector' benchmark, and the whole data sample provided useful understanding and calibration of the combined detector and software performance.

\begin{figure}[ht]
\includegraphics[width=0.49\textwidth]{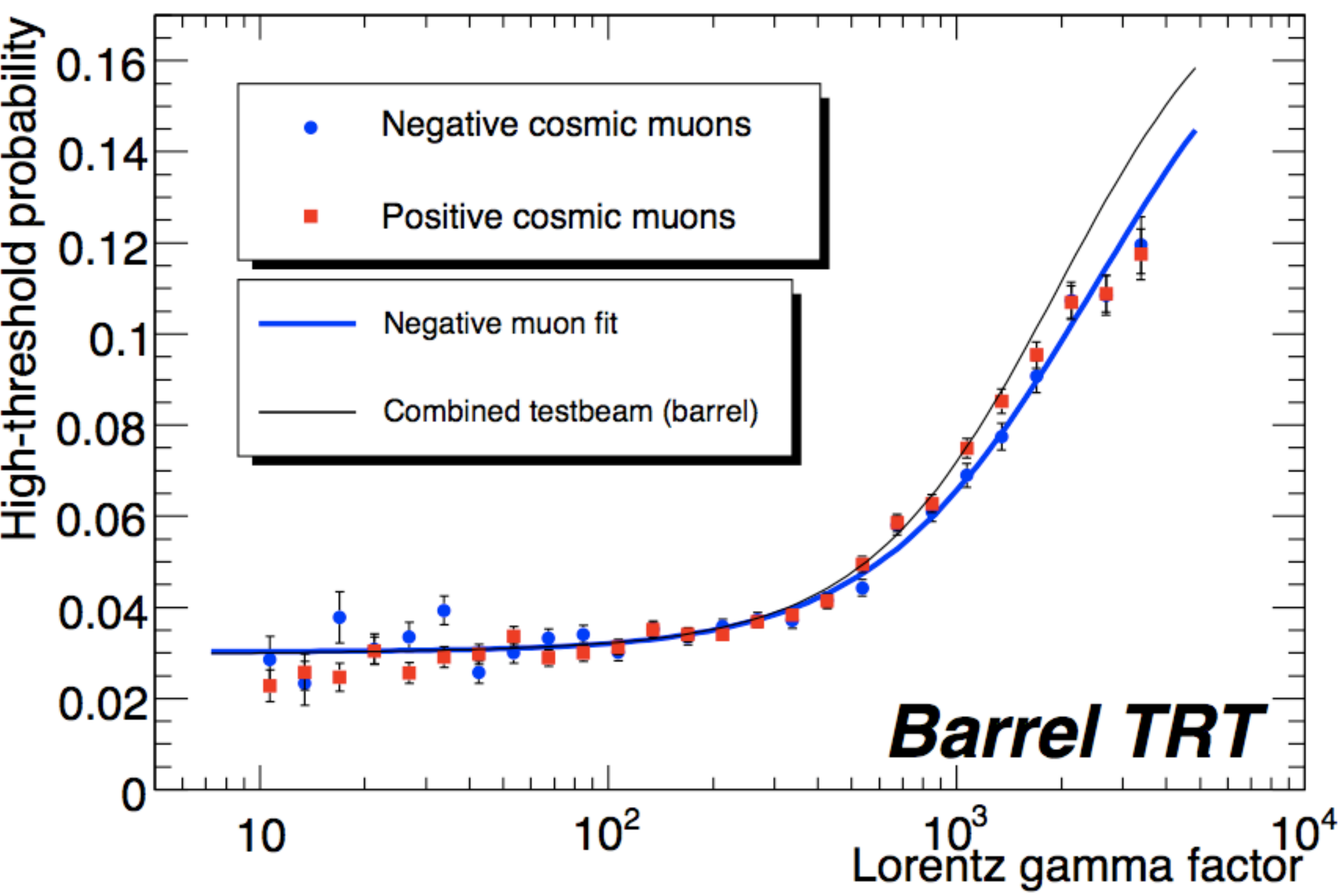}\hspace{0.05\textwidth}%
\begin{minipage}[b]{0.46\textwidth}\caption{\label{fig:cosmics_TRT}Production of transition-radiation photons as a function of $\gamma$ as measured for muon tracks during cosmic data taking of the ATLAS TRT in October~2008. The data points, shown for both muon charges (positive: red squares, negative: blue circles), are fitted (thick blue line) and are compared to the results obtained in the ATLAS Combined Test Beam in 2004 (thin black line). \\}
\end{minipage}
\end{figure}

An example of the performance evaluation during the cosmics run can be seen in~figure~\ref{fig:cosmics_TRT} for the ATLAS Transition Radiation Tracker (TRT). The turn-on of the transition-radiation X-ray emission is nicely seen and the identical behaviour of the detector to cosmic tracks and data recorded at the test beam demonstrates that the TRT is working properly. The correspondence between $\mu^+$ and $\mu^-$ is very good, and the results achieved with the barrel TRT and the endcap TRT also look much alike.

\begin{figure}[ht]
\begin{minipage}[b]{0.42\textwidth}
\includegraphics[width=\textwidth]{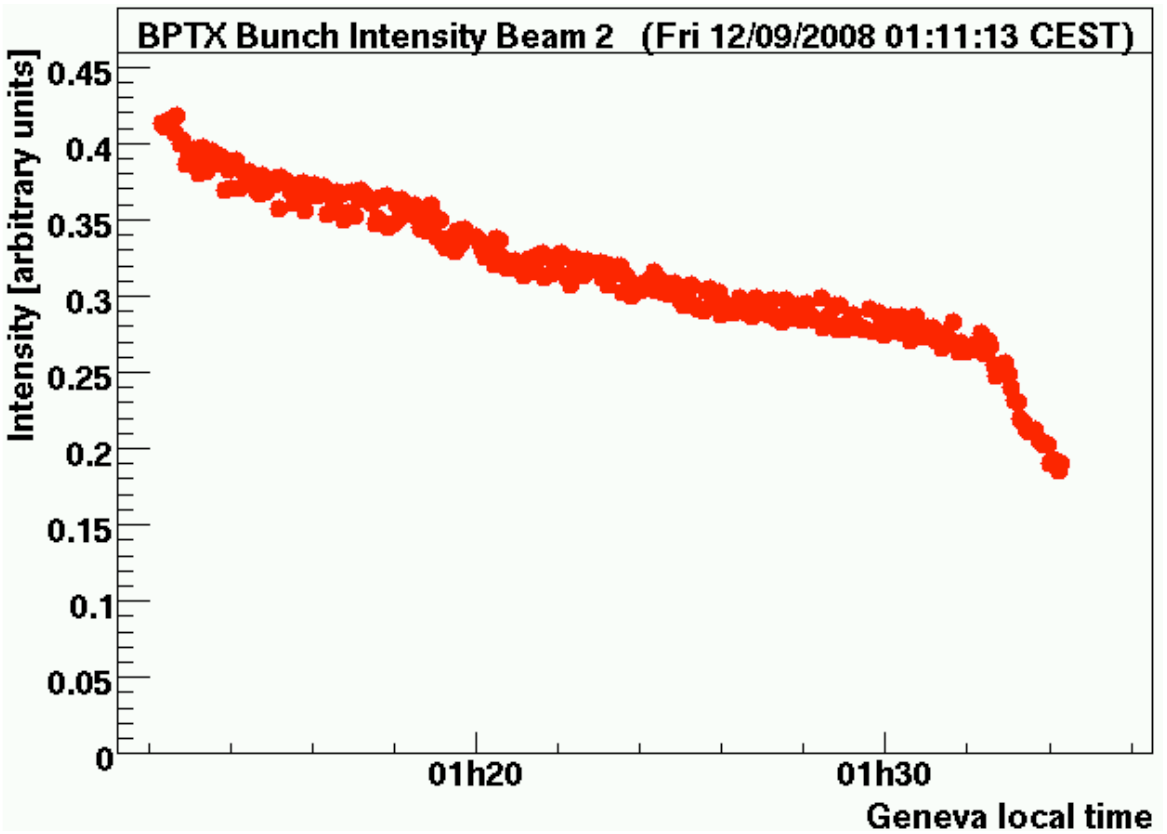}
\caption{\label{fig:beam}Bunch intensity measured by the beam pick-up monitoring system  on September $\rm12^{th}$, 2008.}
\end{minipage}\hspace{0.05\textwidth}%
\begin{minipage}[b]{0.52\textwidth}
\includegraphics[width=\textwidth]{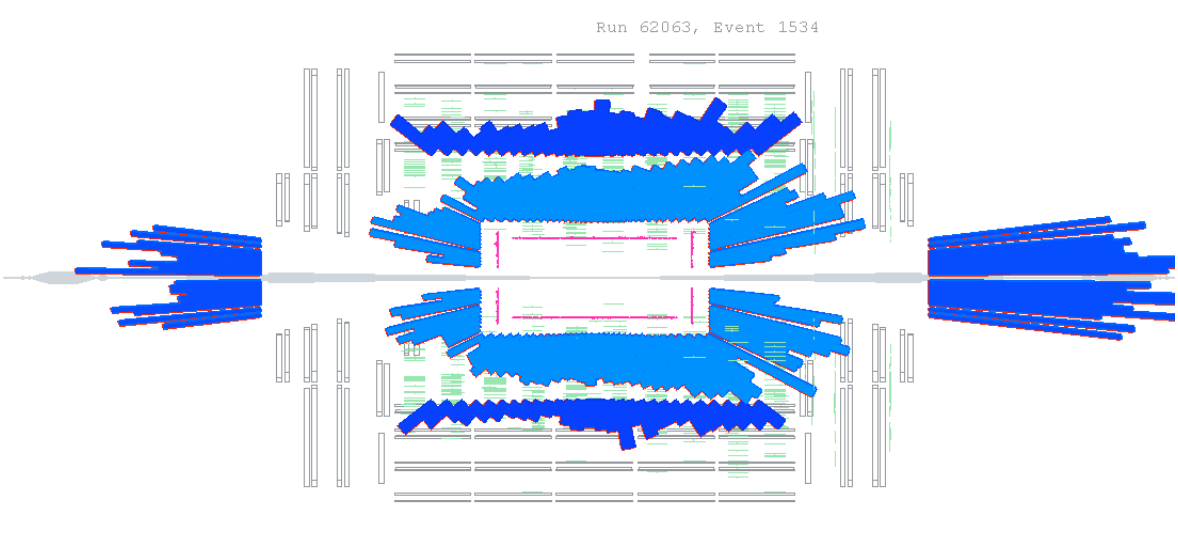}
\caption{\label{fig:CMS_HCAL}Longitudinal view of the energy deposited in the CMS HCAL calorimeter in a splash event.}
\end{minipage} 
\end{figure}

On September $\rm10^{th}$, 2008, and later on for few days, LHC have successfully circulated the first proton beams with a single bunch. For safety considerations, ATLAS has taken the data with Pixel detector was switched off, and some other subsystems were operated with reduced voltage. The solenoid magnet was off, but the toroid systems were operational. The beam intensity recorded on September $\rm12^{th}$, 2008 during a coast of more than 20~minutes of beam~2 is shown in figure~\ref{fig:beam}. The relative precision determined from the scatter of data points is 10\%. The absolute intensity value is not calibrated yet and corresponds roughly to unit of $10^{10}$~protons. In figure~\ref{fig:CMS_HCAL}, a so-called `splash' event is displayed, as debris of particles hitting the collimator blocks around 150~m from LHC Point~5 deposit energy on the CMS hadronic calorimeter. 

More details on the commissioning efforts of ATLAS and CMS can be found in references~\cite{ATLAS-comm} and~\cite{CMS-comm}, respectively, and in references therein.

\section{Early physics: a few examples}%%%%%%%%%%%%%%%%%%%%%%%%%%%%%%%%%%%%%%%%%%%%%%%%%

\begin{figure}[ht]
\begin{minipage}[b]{0.47\textwidth}
\includegraphics[width=\textwidth]{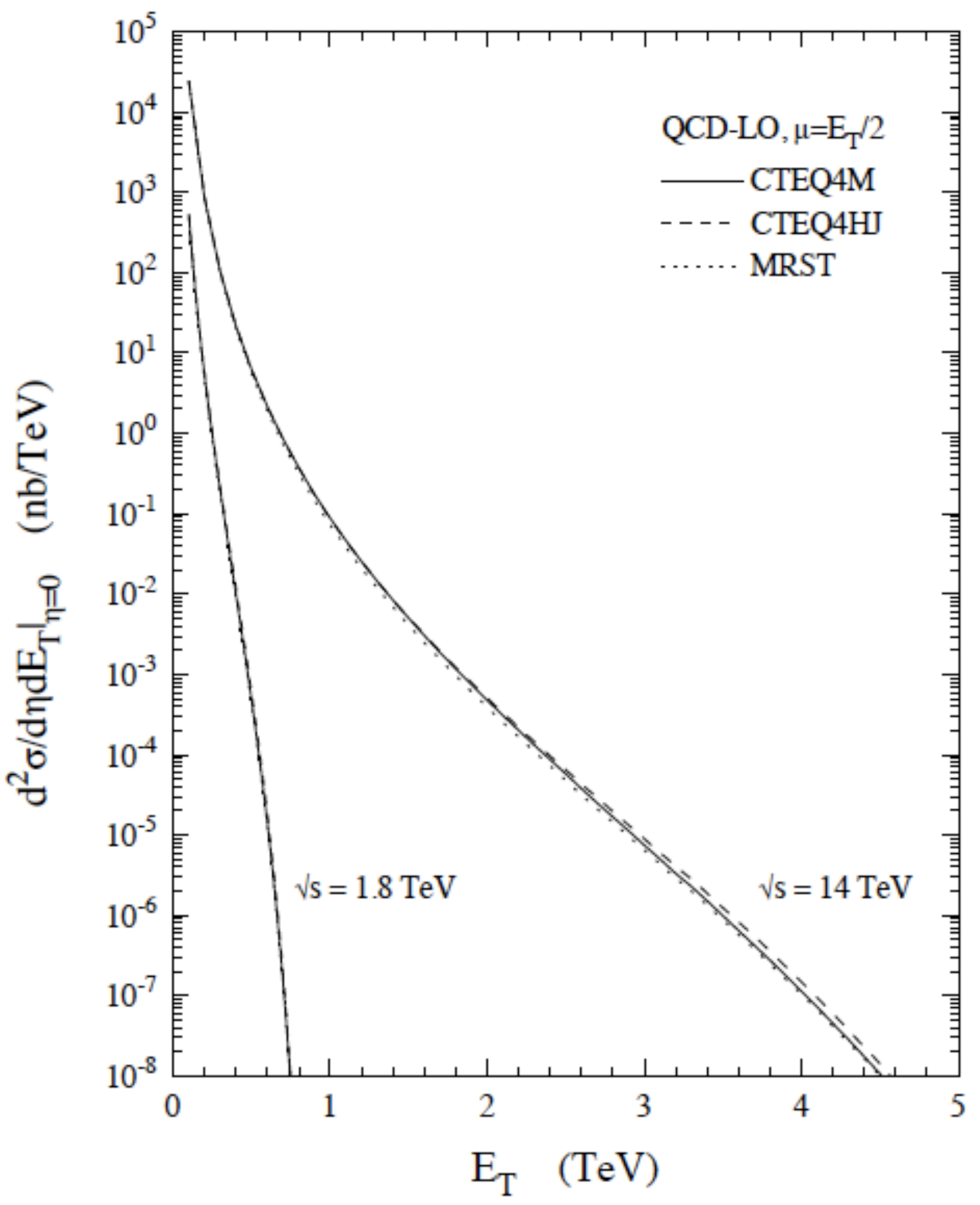}
\caption{\label{fig:LHCxsection}The inclusive jet cross sections at a rapidity of $\eta=0$ for both the Tevatron and the LHC~\cite{LHCxsection}.}
\end{minipage}\hspace{0.05\textwidth}%
\begin{minipage}[b]{0.47\textwidth}
\includegraphics[width=\textwidth]{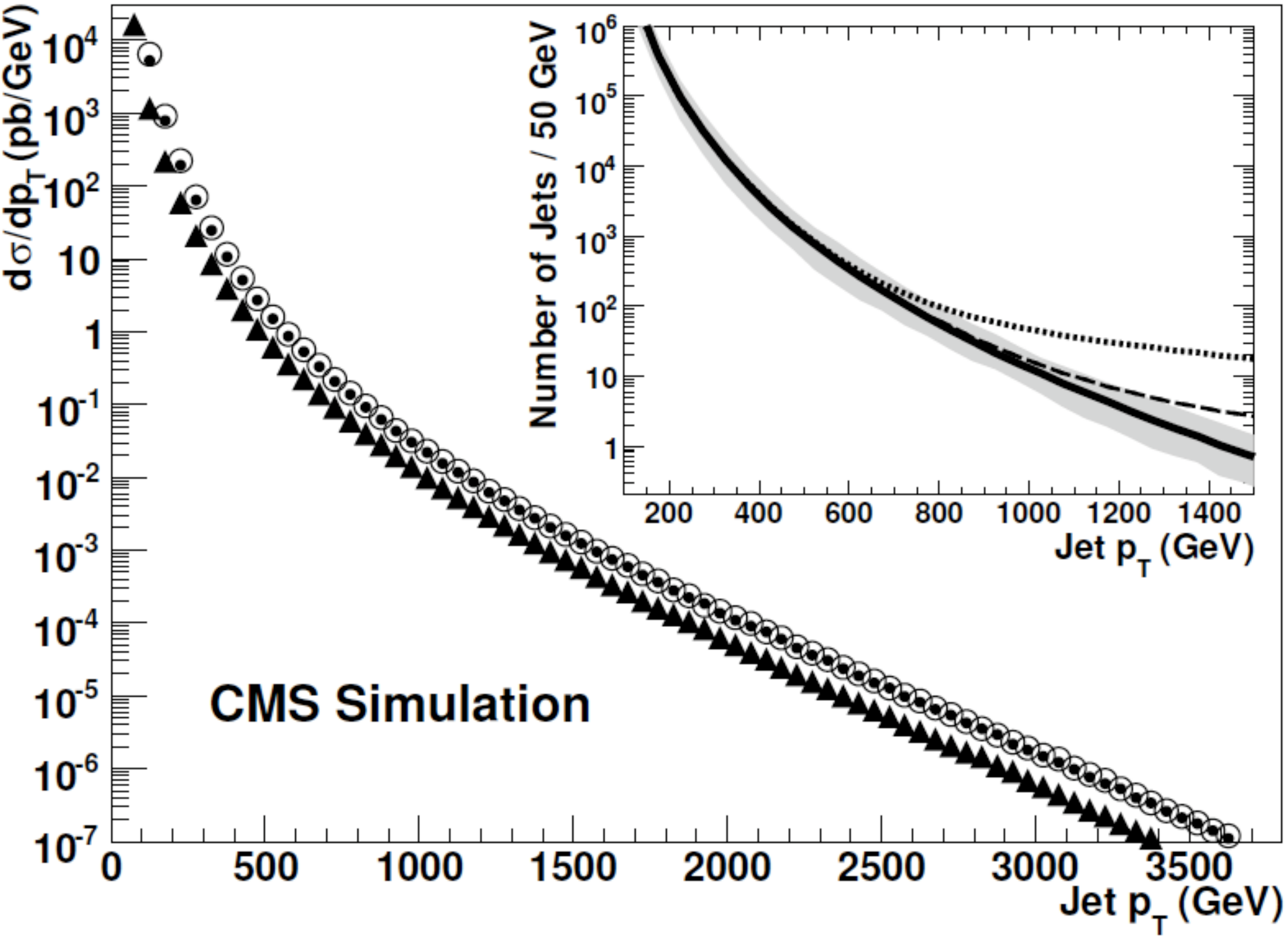}
\caption{\label{fig:contact}The inclusive jet $p_{\rm T}$ differential cross section expected from QCD, for generated jets (points), reconstructed jets (triangles), and corrected jets (open circles). The inset shows the number of generated jets expected for $\rm10~pb^{-1}$. The standard QCD curve (solid) is modified by a signal from contact interactions with scale $\Lambda^+ = 3~{\rm TeV}$ (dotted) and 5~TeV (dashed). The shaded band represents the effect of a 10\% uncertainty on the jet energy scale.~\cite{CMS-contact}.}
\end{minipage} 
\end{figure}

As the two LHC experiments start accumulating collision data, it will be possible to probe various aspects of Standard Model and explore new physics scenarios. According to the current schedule of the machine, shown in table~\ref{tab:operation}, the first delivered luminosity will be around $\rm10^{30}~cm^{-2}s^{-1}$. Even with a few tens of pb$^{-1}$ expected by the end of 2009, many interesting results can be obtained as will be shown in the following. The main point is that the ${\cal O}(10~{\rm TeV})$ energy regime is going to be unchartered territory for particle physics. This is clearly  demonstrated in figure~\ref{fig:LHCxsection}, where the inclusive jet cross section for the LHC ($\sqrt{s}=14$~TeV) is compared to that of the Tevatron Run~I ($\sqrt{s}=1.8$~TeV). However, early data analysis will focus mostly on SM processes with two goals: (i) understanding the performance of complex detectors such as ATLAS and CMS; and (ii) measuring basic SM processes and comparing to theory and various Monte Carlo tools. The expected physics performance for the ATLAS and CMS experiments has been documented in detail in references~\cite{ATLAS-CSC} and~\cite{CMS-TDR}, respectively.

\begin{table}[ht]
\caption{\label{tab:operation}The expected LHC operation parameters in terms of centre-of-mass energy and collected integrated luminosity (as of December~2008).}
\begin{center}
\lineup
\begin{tabular}{lllll}
\br
Year: & 2009 & 2010 & 2010--2012 & $>2012$ \\
\mr
$\int L{\rm d}t$/yr & 10--100~$\rm pb^{-1}$ & 0.5--2~fb$^{-1}$ & $\cal{O}$(10~fb$^{-1}$)  & $\cal{O}$(100~fb$^{-1}$) \\
Energy & 8--10~TeV & 14~TeV & 14~TeV & 14~TeV \\
\br
\end{tabular}
\end{center}
\end{table}

The available collision data will be completely dominated by minimum bias and QCD jet events, as shown in figure~\ref{fig:LHCxsection}. This will allow the study of the underlying event using minimum bias and di-jet samples and the tuning of the relevant Monte Carlo generators. Besides that, and despite the theoretical uncertainties on the QCD production at these energies, search for new phenomena will be feasible. For example, contact interactions will be accessible during the first year of LHC operation. As demonstrated in figure~\ref{fig:contact}, contact interactions induce large rates on high-$p_{\rm T}$ jets. The experimental error will be dominated by the jet energy scale ($\sim$10\%) during early running, however this will not compromise the discovery potential for $p_{\rm T}>1\,{\rm TeV}$ at $\Lambda^{+}=3$~TeV~\cite{CMS-contact}. This means that even with only 10~pb$^{-1}$ the current bound of $\Lambda^{+}>2.7$~TeV set by Tevatron can be superseded. 

The cross-sections for various physics processes producing isolated leptons at the~LHC are quite large. Taking electrons in~ATLAS as an example, for an integrated luminosity of 100~pb$^{-1}$, one expects to trigger and reconstruct on roughly $250\,000$~$\rm J/\psi$, $10^{6}$~$\rm W\rightarrow e\nu$ and~$50\,000$~$\rm Z\rightarrow ee$ decays. For early data, the single (resp.~double) electron trigger thresholds range from~10 to~20~GeV (resp.~5 to~15~GeV), and the collection of large samples of isolated electrons will be possible rapidly, a crucial element for the understanding of the initial detector and trigger performance.

%\begin{figure}[ht]
%\includegraphics[width=0.63\textwidth]{electrons}\hspace{0.05\textwidth}
%\begin{minipage}[b]{0.32\textwidth}
%\caption{\label{fig:electrons}LHC cross sections for various processes producing electrons as a function of the electron (true) transverse momentum. \\}
%\end{minipage}
%\end{figure}

\begin{figure}[ht]
\begin{minipage}[b]{0.46\textwidth}
\includegraphics[width=\textwidth]{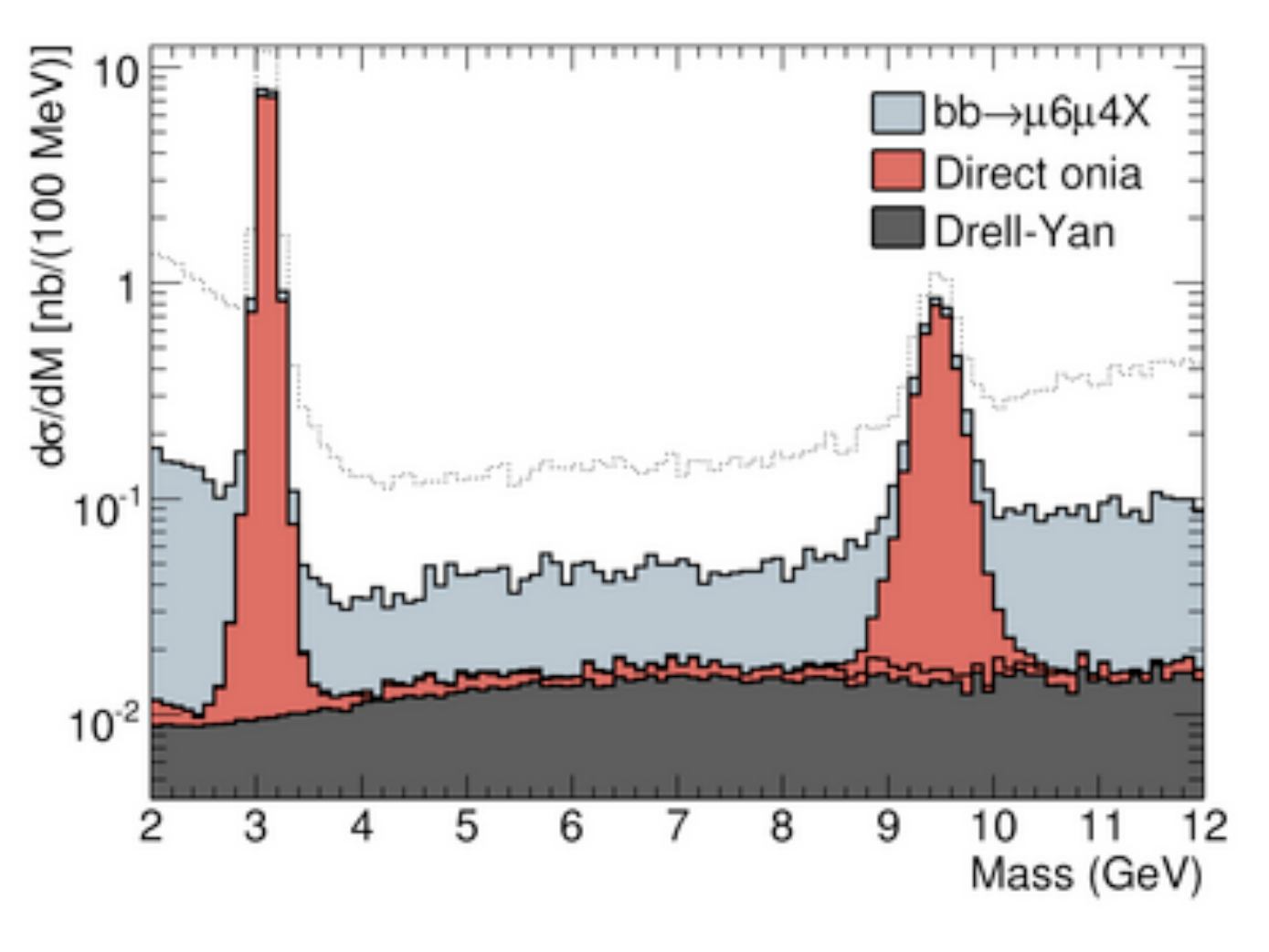}
\caption{\label{fig:ATLAS_muons} Distribution of di-muon invariant mass in the region of the~$\rm J/\psi$ and~$\Upsilon$ resonances in ATLAS, shown above the backgrounds from heavy-flavour and Drell-Yan production.}
\end{minipage}\hspace{0.05\textwidth}%
\begin{minipage}[b]{0.49\textwidth}
\includegraphics[width=\textwidth]{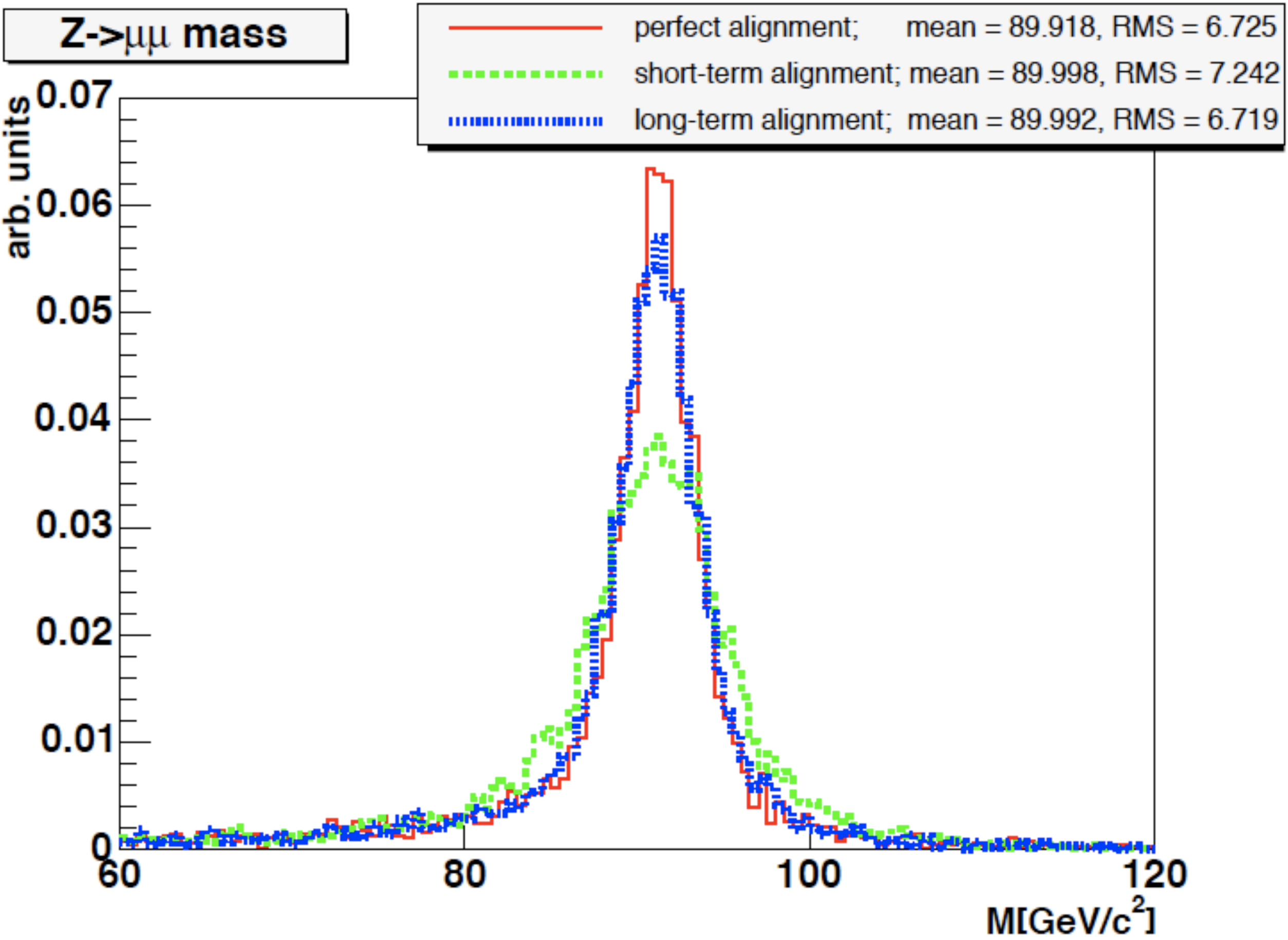}
\caption{\label{fig:Zmumu}Z mass reconstruction from di-muons in CMS for various (mis)alignment scenarios~\cite{Z_mumu}.}
\end{minipage} 
\end{figure}

Similarly to electrons, the detection and measurement of di-muons will provide an excellent opportunity  to collect large samples to understand the performance of the relevant detectors and trigger. The invariant mass of the di-muons for the two most prominent states from low-mass resonances are displayed in~figure~\ref{fig:ATLAS_muons}. After all selection cuts, ATLAS will record about~4200 (resp.~800) events from~$\rm J/\psi$ (resp.~$\Upsilon$) decays to a muon pair 
per day at a luminosity of~$\rm 10^{31}~cm^{-2}s^{-1}$, assuming an overall 30\%~machine and detector data-taking efficiency. These events will provide quantitative information on the tracker momentum scale, on the trigger performance, the individual detector efficiencies, etc. However, care will have to be devoted to minimise and control the backgrounds from heavy flavours and $\pi$/K~decays.

High-mass resonances such as the Z peak will also provide crucial information. After all cuts, about 160 $\rm Z\rightarrow\mu\mu$ events per day are expected to be observed with a luminosity of $L=\rm 10^{31}~cm^{-2}s^{-1}$. This corresponds to a rate for Z events 10 times higher than the Tevatron. The precision on the $\rm Z\rightarrow\mu\mu$ cross section measurement is limited with 100~$\rm pb^{-1}$ by $<2\%$ experimental errors, such as mis-alignment of the muon spectrometer (see figure~\ref{fig:Zmumu}) and by $\sim10\%$ by the luminosity error. These studies are expected to aid at the muon spectrometer alignment and at the determination of the EM calorimeter uniformity and the energy/momentum scale of the full detector. The intercalibration of the EM and the hadronic calorimeter using azimuthal symmetry is also possible. Additionally, the lepton trigger and reconstruction efficiency will be better understood.

\begin{figure}[ht]
\begin{minipage}[b]{0.5\textwidth}
\includegraphics[width=\textwidth]{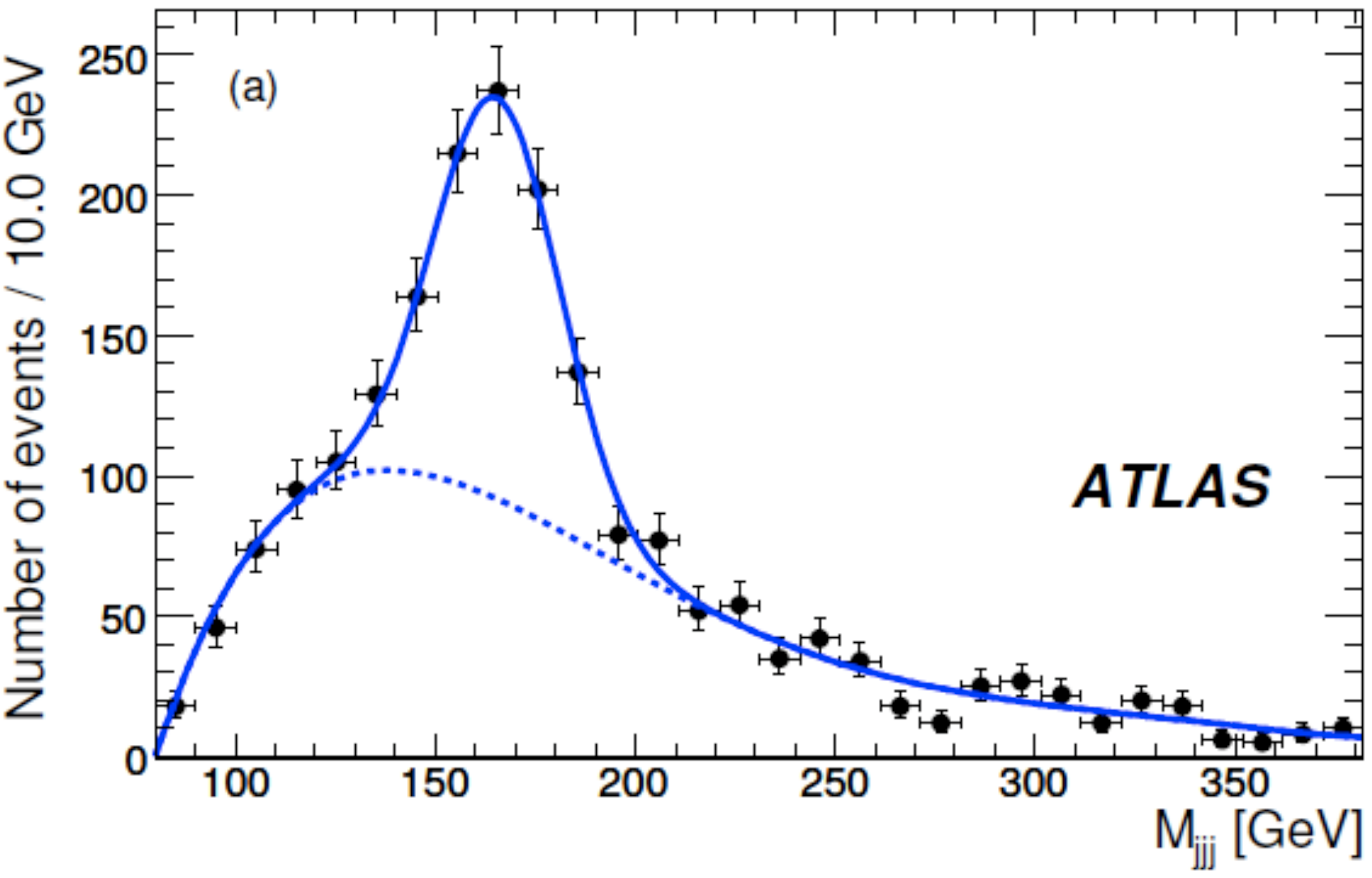}
\caption{\label{fig:top_ATLAS}Fit to the three-jet invariant mass in ATLAS for $\rm 200~pb^{-1}$: the Chebychev polynomial fit to the background (dotted line) and the Gaussian fit of the signal events (full line) are shown~\cite{ATLAS-CSC}.}
\end{minipage}\hspace{0.05\textwidth}%
\begin{minipage}[b]{0.45\textwidth}
\includegraphics[width=\textwidth]{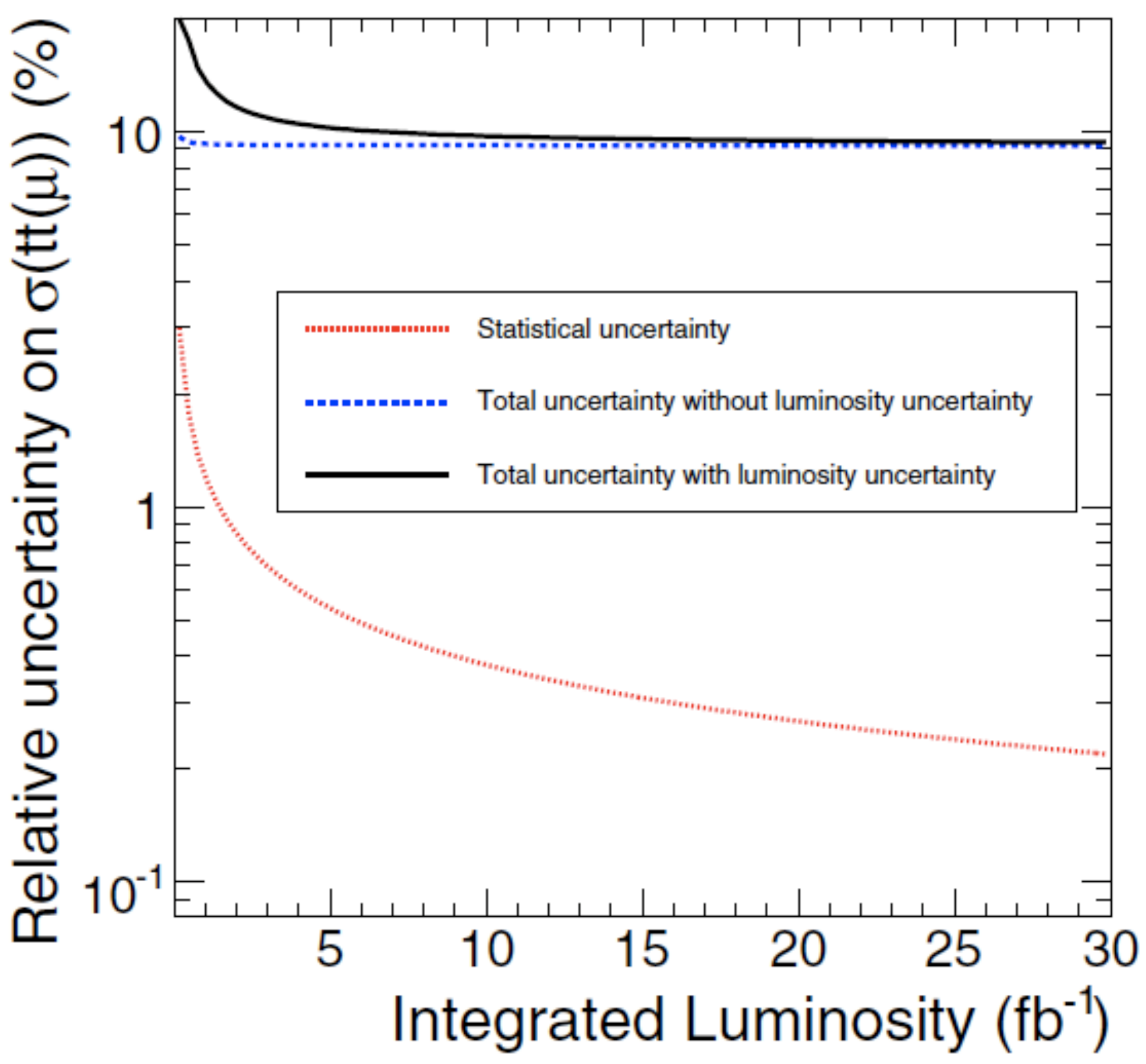}
\caption{\label{fig:top_CMS}Statistical and total uncertainty on the inferred cross section by CMS of the process $\rm pp\rightarrow t\bar{t}\rightarrow bq\bar{q}\mu\nu_{\mu}$ versus the integrated luminosity~\cite{CMS-TDR}.}
\end{minipage} 
\end{figure}

The first top quarks outside Fermilab will be seen at the LHC. The top-pair signal in the semi-leptonic channel can be extracted using event counting or the three-jet invariant distribution. The latter is demonstrated in figure~\ref{fig:top_ATLAS}. A signal can be established with 100~$\rm pb^{-1}$ even if a pessimistic background knowledge is assumed. With this integrated luminosity, the b-tagging is expected to be understood and the its efficiency known to $<5\%$, allowing thus its use in the event selection. By requiring one or two b-jets, the non-$\rm t\bar{t}$ background will be reduced and the selection of the correct combination will be improved. As seen in figure~\ref{fig:top_CMS}, with ${\cal O}(1~{\rm fb}^{-1})$ the b-tagging and the knowledge of the parton distribution functions are important, whist at higher integrated luminosities the error on the luminosity dominates the overall uncertainty.

\begin{figure}[ht]
\includegraphics[width=0.47\textwidth]{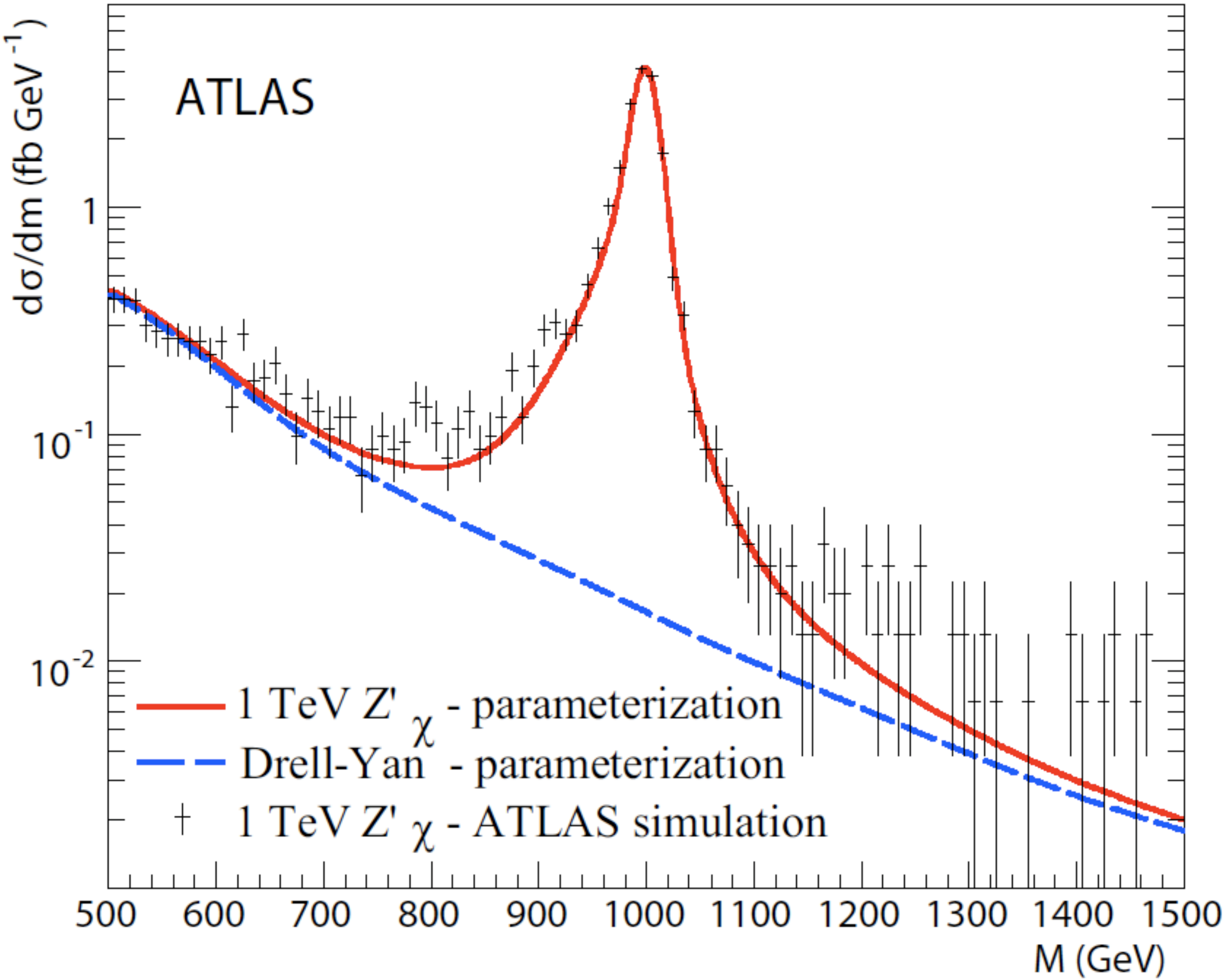}\hspace{0.05\textwidth}%
\begin{minipage}[b]{0.48\textwidth}\caption{\label{fig:Z_prime}Mass spectrum for a $m = 1$~TeV $\rm Z'_{\chi}\rightarrow e^+e^-$ obtained with ATLAS full simulation (histogram)
and the parameterisation (solid line). The dashed line corresponds to the parameterisation of the Drell-Yan process (irreducible background) \cite{ATLAS-CSC}.\\}
\end{minipage}
\end{figure}

Obvious candidates for first searches are heavy resonances decaying to leptons. A number of benchmark $\rm Z'$ models has been studied at LHC: the Sequential Standard Model ($\rm Z'_{SSM}$), the $E_6$ models $\rm Z'_{\psi}$, $\rm Z'_{\chi}$, $\rm Z'_{\eta}$, and the left-right symmetric model ($\rm Z'_{LR}$). With 100~$\rm pb^{-1}$ a signal large enough for a $5\sigma$ discovery can be observed for masses up to 1~TeV, as seen in table~\ref{tab:resonance}. Indicatively, the Tevatron reach for $5\sigma$ discovery with 7~$\rm fb^{-1}$ of collected data is $\sim 1$~TeV. The ultimate ATLAS reach, with 300~$\rm fb^{-1}$ is expected to be $\sim5$~TeV. The signal can be seen as a (narrow) mass peak on top of the small (at these energies) Drell-Yan background. This is demonstrated in figure~\ref{fig:Z_prime} for a $\rm Z'_{\chi}$ with a mass of 1~TeV. The ultimate calorimeter performance is not required for these studies, rendering them thus independent of the calorimeter calibration procedures. The di-tau signature is another possibility for the search for high mass resonances. This becomes very important, in particular, in models where a hypothetical new resonance couples preferentially to the third generation~\cite{ATLAS-CSC}.

\begin{table}[ht]
\caption{\label{tab:resonance}Event rate after all analysis cuts and integrated luminosity required for $5\sigma$ discovery (corresponding to 10~observed events) for $\rm Z'_{SSM}$ decaying to leptons. } %{\bf [citation needed]}}
\begin{center}
\lineup
\begin{tabular}{lll}
\br
Mass [TeV] & Events / $\rm fb^{-1}$ & $\int L{\rm d}t$ [pb$^{-1}$]\\
\mr
1.0 & $\sim160$    & \0\0$\sim70$ \\
1.5 & \0$\sim30$   & \0$\sim300$ \\
2.0 & \0\0$\sim7$  & $\sim1500$ \\
\br
\end{tabular}
\end{center}
\end{table}

\begin{figure}[ht]
\begin{minipage}[b]{0.42\textwidth}
\includegraphics[width=\textwidth]{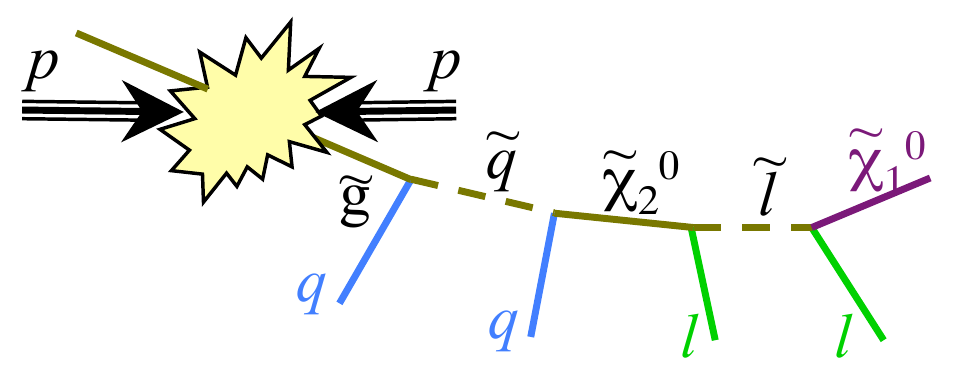}
\caption{\label{fig:SUSY_cascade}Schematic view of a typical supersymmetric cascade decay at the LHC. The $\rm\tilde{q}\tilde{g}$, $\rm\tilde{g}\tilde{g}$ and $\rm\tilde{q}\tilde{q}$ production processes dominate, whilst the gaugino pair production is suppressed. The other $\rm\tilde{g}$ or $\rm\tilde{q}$ decays in a similar manner (not shown).}
\end{minipage}\hspace{0.05\textwidth}%
\begin{minipage}[b]{0.53\textwidth}
\includegraphics[width=\textwidth]{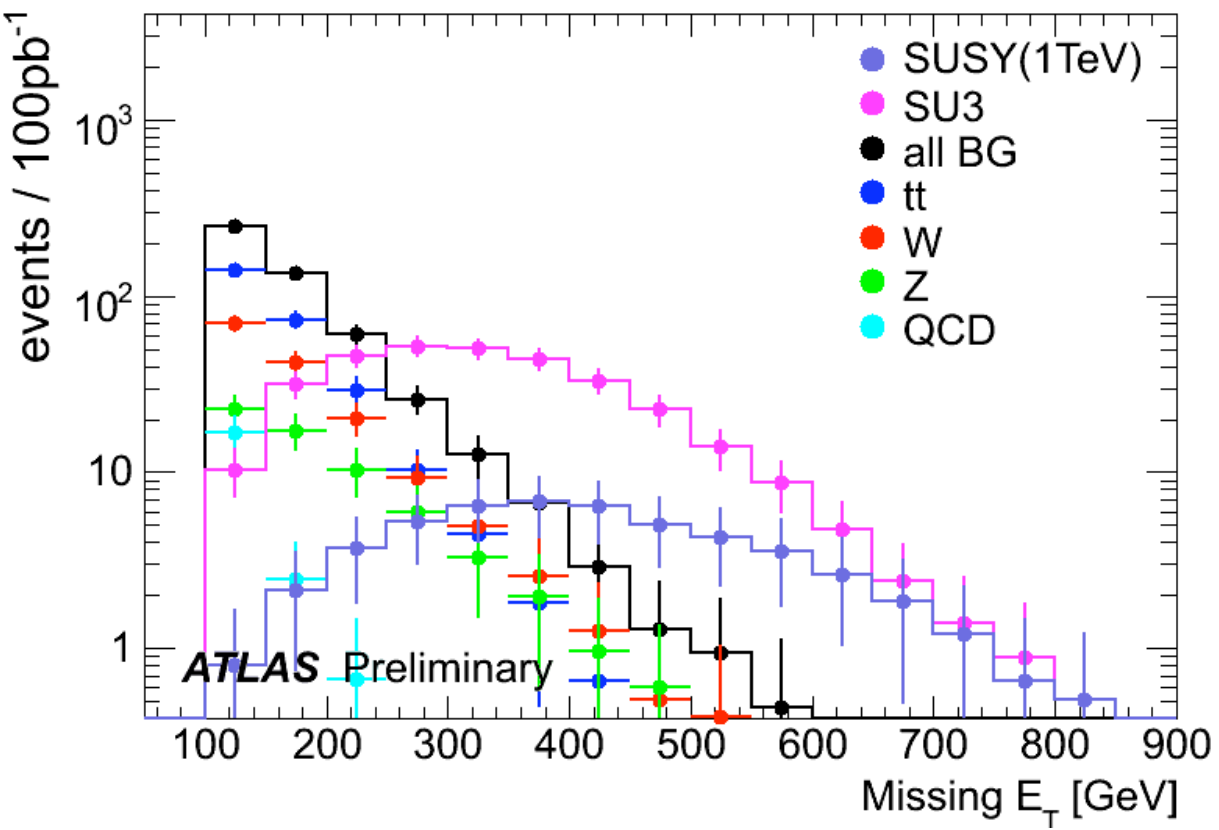}
\caption{\label{fig:SUSY_MET} $E{\rm _{T}^{miss}}$ distributions for various background processes, for the ATLAS SUSY point SU3 (700~GeV) and for SUSY at 1-TeV scale in the no-lepton mode for an integrated luminosity of 1~fb$^{-1}$~\cite{ATLAS-CSC}.}
\end{minipage} 
\end{figure}

Furthermore, the first LHC data will allow to perform searches for $R$-parity conserving supersymmetry (SUSY). If it exists at the TeV scale, it should be found `quickly' exploiting the large (strong) production cross sections for
$\rm\tilde{q}\tilde{g}$, $\rm\tilde{g}\tilde{g}$ and $\rm\tilde{q}\tilde{q}$. Spectacular signatures, involving many jets, leptons and $E{\rm _{T}^{miss}}$ ---due to the (stable) lightest neutralinos escaping the detector without interacting with it---are expected from the cascade decay of superpartners, as shown in figure~
\ref{fig:SUSY_cascade}. At the LHC, for $m_{\rm\tilde{q}},\,m_{\rm\tilde{g}}\approx1$~TeV, about 10~events/day are expected at $L=10^{32}~{\rm cm^{-2}s^{-1}}$.
Hints for SUSY up to 1~TeV can be seen at the LHC with only 100~$\rm pb^{-1}$, whilst the 95\% C.L.\ reach at the Tevatron is up to $\sim400$~GeV. The analysis will be based to a large extent to the high missing energy that characterises the SUSY events, as demonstrated in figure~\ref{fig:SUSY_MET}. However precise understanding of the relevant backgrounds will require $\rm\sim1~fb^{-1}$. To this respect, $R$-parity \emph{violating} SUSY may be easier to be probed, since it does not rely on 
$E{\rm _{T}^{miss}}$-based selection criteria. In order to bypass the uncertainty on the missing-energy measurement during the first data taking, various techniques providing data-driven background estimation have been developed~\cite{ATLAS-CSC} for SUSY searches.

\begin{figure}[ht]
\begin{center}
\includegraphics[width=0.7\textwidth]{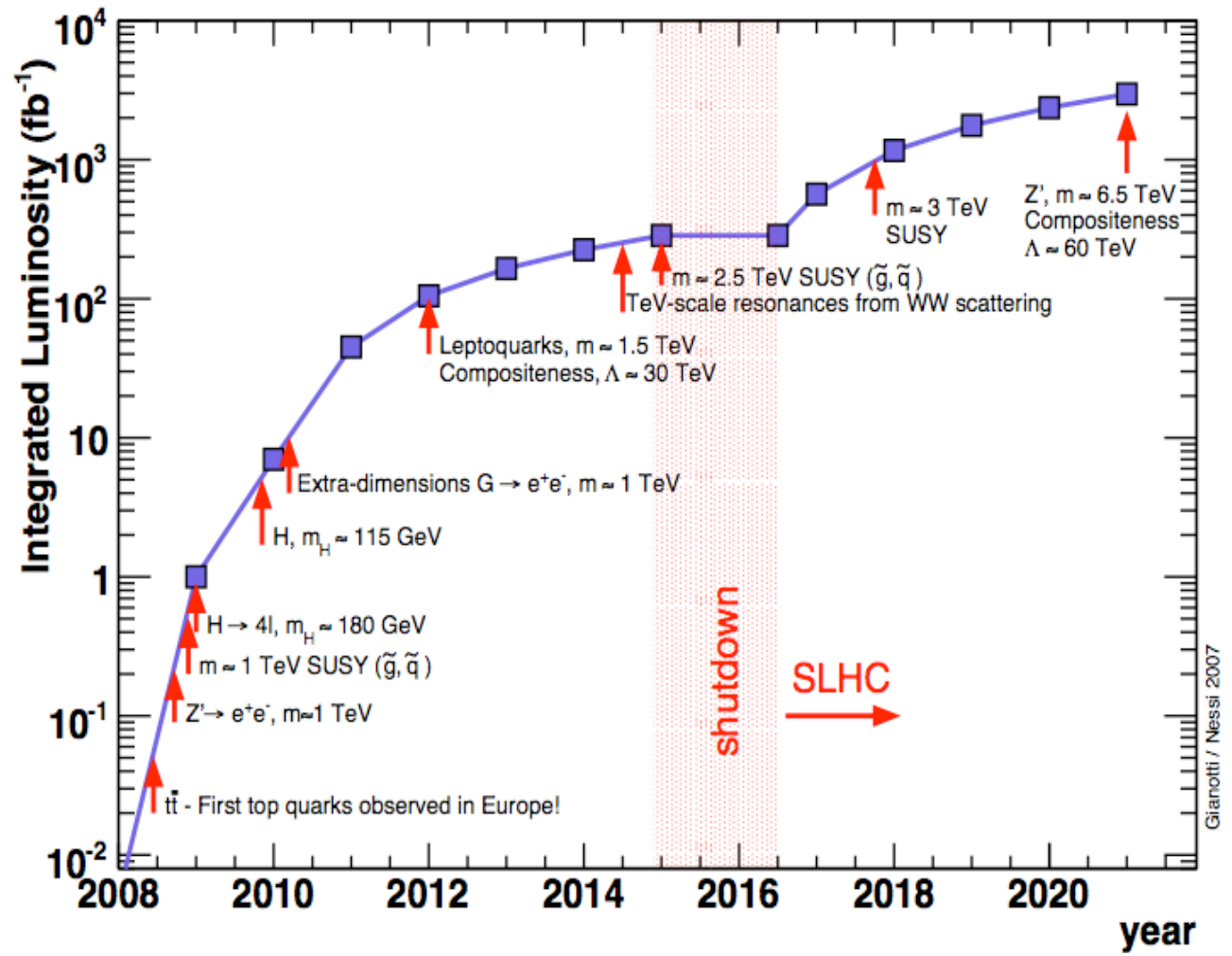}
\end{center}
\caption{\label{fig:physics_program}LHC discovery potential versus time and integrated luminosity (Credit: F.~Gianotti and M.~Nessi, 2007).}
\end{figure}

The eventual discovery reach of the LHC as a function of the collected integrated luminosity (and time) is summarised in figure~\ref{fig:physics_program}. A sample of exotic physics scenarios that will be explored at the LHC in the next few years is given below.
\begin{description}
\item[Excited quarks] They are predicted in composite-quark models. Through their decay to photons, $\rm q^*\rightarrow q\gamma$, they can be probed up to  $m\approx6$~TeV, if the compositeness scale is less than the LHC energy.
\item[Leptoquarks] The experimentally observed symmetry between leptons and quarks has motivated the search for leptoquarks, hypothetical bosons carrying both baryon
and lepton quantum numbers. Their discovery will be feasible through their pair production up to $m\approx1.5$~TeV.
\item[Monopoles] The existence of Dirac magnetic monopoles will be investigated at the LHC via the production of a photon pair from a virtual monopole loop, $\rm pp\rightarrow pp\gamma\gamma$, for masses $m\lesssim20$~TeV.
\item[Compositeness] If the compositeness scale $\Lambda$ is much larger than the centre of mass energy of the colliding partons, $\sqrt{s}=14~{\rm TeV}$, the manifestation of compositeness will be an effective 4-fermion contact interaction. The angular distributions and rates of high-$p_{\rm T}$ di-jets and di-muons will allow to probe fermion compositeness up to $\Lambda\approx40$~TeV.
\item[Heavy resonances] New heavy states forming a narrow resonance decaying into opposite sign di-leptons are predicted in many extensions of the Standard Model: grand unified theories, technicolor, little Higgs models, and models involving extra dimensions. Di-lepton and di-jet states, such as $\rm Z'\rightarrow\ell\ell,\,qq$, will be probed  up to $m\approx5$~TeV, whereas resonances decaying to a lepton and a neutrino ($\rm W'\rightarrow\ell\nu$) will be searched for masses $m\lesssim6$~TeV.
%\item[Technicolor] Dynamical electroweak symmetry breaking can be induced by techni-fermions, which interact strongly at high scale. Searches for techni-pions are feasible, e.g.\ through the channel $\rm\phi_{T}\rightarrow WZ\rightarrow\ell\nu\ell\ell$ for masses up to 
%\item[Extra dimensions]Ê
\end{description}

The opportunities for the discovery of new physics at the~LHC are numerous and the~ATLAS and CMS~detectors have been optimised very systematically towards maximum sensitivity to almost any new physics signals, based on detailed benchmarks from many models. Nevertheless, even some known physics measurements might be beyond the reach of the~LHC (and even of the~SLHC), such as the Higgs-boson self-coupling and the detection of charginos and neutralinos in many supersymmetry scenarios. 

\section{Outlook}%%%%%%%%%%%%%%%%%%%%%%%%%%%%%%%%%%%%%%%%%%%%%%%%%%%%%%%%%%%%%%%%%%%%%%

The LHC accelerator and its experiments represent the front wave of the field of experimental particle physics and it is a wonderful achievement of the community across the whole world that these very challenging projects have all been able to successfully conclude their last phases of installation and commissioning in~2008. These projects have been with us however for almost twenty years and have become gigantic in terms of their complexity, their funding and their human resources. One could argue that the whole project has become dinosaur-like in terms of sheer size and scope.

The most prominent issue with the scope of the LHC experiments is related to the very long time-scales involved which threaten to breach the continuity ensured between generations of experimental and theoretical physicists through rapid turnover between new ideas and concepts, both in theory and technical developments, new experiments, and new physics results, leading in turn to new ideas and concepts. Theory and foremost among all the Standard Model has not been sufficiently challenged nor nourished by experiment for too long (leaving aside in this context the exciting results obtained in neutrino physics over the past ten-fifteen years). Most of the young physicists now joining~ATLAS and~CMS have no idea of what really has been installed in the experimental halls, nor of how formidable a challenge it has been to build, install and commission these detectors. Very few people remember today that in the late 1980's most experimental physicists believed it would be impossible to operate tracking detectors in the radiation environment of the LHC~accelerator.

The stakes of these projects have therefore become very high and it is out of the question nowadays to afford shots in the dark of such a large size and scope. Success must be guaranteed, if not in terms of physics results, at least in terms of achieved technical goals and detector performance for known physics processes. This also probably means that one can no longer afford, as in the early 1980's, to approve the next large machine before the current one has delivered some results.

On the other hand, this is why the challenge in front of us at the~LHC is so exhilarating! A major fraction of the future of our discipline hangs on the physics which will be harvested at this new energy frontier. How ordinary or extraordinary will this harvest be? Only nature knows.

\end{document}